\documentclass[namedreferences]{solarphysics}
\usepackage[hyperref,optionalrh,solaromanenum]{spr-sola-addons} 
\usepackage{graphicx}                    
\usepackage{color}                       
\usepackage{breakurl}                         

\def\arcsec{\hbox{$^{\prime\prime}$}}
\def\arcmin{\hbox{$^{\prime}$}}
\def\degns{\ifmmode^\circ\else$^\circ$\fi}
\def\deg{\ifmmode^\circ\else$^\circ$\fi}
\def\uv{{\sl uv}}

\def\eg{e.g.,~}
\def\ie{i.e.,~}

\def\gsim{\lower.4ex\hbox{$\;\buildrel >\over{\scriptstyle\sim}\;$}}
\def\lsim{\lower.4ex\hbox{$\;\buildrel <\over{\scriptstyle\sim}\;$}}

\newcommand{\aap}{    {\it Astron. Astrophys.}}

\newcommand{\araa}{   {\it Ann. Rev. Astron. Astrophys.}}

\newcommand{\apj}{    {\it Astrophys. J.}}

\newcommand{\pasp}{   {\it Pub. Astron. Soc. Pac.}}
\newcommand{\pasj}{   {\it Pub. Astron. Soc. Japan}}

\newcommand{\solphys}{{\it Solar Phys.}}
 
\newcommand{\ssr}{    {\it Space Sci. Rev.}} 
\newcommand{\procspie}{{\it Proceedings of the SPIE}}
\newcommand{\mdash}{{}}
\chardef\us=`\_

\begin{document}

\begin{article}

\begin{opening}

\title{Observing the Sun with Atacama Large Millimeter/submillimeter Array (ALMA): High Resolution Interferometric Imaging}

%
\author[addressref={1,2},corref,email={masumi.shimojo@nao.ac.jp}]{\inits{M.}\fnm{M.}~\lnm{Shimojo}\orcid{0000-0002-2350-3749}}
\author[addressref={3}]{\inits{T.S.}\fnm{T.S.}~\lnm{Bastian}}
\author[addressref={4,3}]{\inits{A.S.}\fnm{A.S.}~\lnm{Hales}}
\author[addressref={5}]{\inits{S.M.}\fnm{S.M.}~\lnm{White}}
\author[addressref={6,7}]{\inits{K.}\fnm{K.}~\lnm{Iwai}}
\author[addressref={8}]{\inits{R.E.}\fnm{R.E.}~\lnm{Hills}}
\author[addressref={4,1}]{\inits{A.}\fnm{A.}~\lnm{Hirota}}
\author[addressref={4,9}]{\inits{N.M.}\fnm{N.M.}~\lnm{Phillips}}
\author[addressref={4,1}]{\inits{T.}\fnm{T.}~\lnm{Sawada}}
\author[addressref={10}]{\inits{P.}\fnm{P.}~\lnm{Yagoubov}}
\author[addressref={4,9}]{\inits{G.}\fnm{G.}~\lnm{Siringo}}
\author[addressref={1}]{\inits{S.}\fnm{S.}~\lnm{Asayama}}
\author[addressref={1}]{\inits{M.}\fnm{M.}~\lnm{Sugimoto}}
\author[addressref={11}]{\inits{R.}\fnm{R.}~\lnm{Braj\v{s}a}}
\author[addressref={12}]{\inits{I.}\fnm{I.}~\lnm{Skoki\'{c}}}
\author[addressref={12}]{\inits{M.}\fnm{M.}~\lnm{B\'{a}rta}}
\author[addressref={13}]{\inits{S.}\fnm{S.}~\lnm{Kim}}
\author[addressref={4,9}]{\inits{I.}\fnm{I.}~\lnm{de Gregorio}}
\author[addressref={4,3}]{\inits{S.A.}\fnm{S.A.}~\lnm{Corder}}
\author[addressref={14,15}]{\inits{H.S.}\fnm{H.S.}~\lnm{Hudson}}
\author[addressref={16}]{\inits{S.}\fnm{S.}~\lnm{Wedemeyer}}
\author[addressref={17}]{\inits{D.E.}\fnm{D.E.}~\lnm{Gary}}
\author[addressref={18,16}]{\inits{B.}\fnm{B.}~\lnm{De Pontieu}}
\author[addressref={17,19,20}]{\inits{M.}\fnm{M.}~\lnm{Loukitcheva}}
\author[addressref={17}]{\inits{G.D.}\fnm{G.D.}~\lnm{Fleishman}}
\author[addressref={17}]{\inits{B.}\fnm{B.}~\lnm{Chen}}
\author[addressref={21}]{\inits{A.}\fnm{A.}~\lnm{Kobelski}}
\author[addressref={22}]{\inits{Y.}\fnm{Y.}~\lnm{Yan}}

\address[id=1]{National Astronomical Observatory of Japan (NAOJ), 2-21-1, Osawa, Mitaka, Tokyo 181-8588, Japan}
\address[id=2]{Department of Astronomical Science, The Graduate University for Advanced Studies (SOKENDAI), 2-21-1, Osawa, Mitaka, Tokyo 181-8588, Japan}
\address[id=3]{National Radio Astronomy Observatory (NRAO), 520 Edgemont Road, Charlottesville, VA 22903, USA}
\address[id=4]{Joint ALMA Observatory (JAO), Alonso de C\'{o}rdova 3107, Vitacura 763-0355, Santiago, Chile}
\address[id=5]{Space Vehicles Directorate, Air Force Research Laboratory, 3550 Aberdeen Avenue SE, Kirtland AFB, NM 87117-5776, USA}
\address[id=6]{Institute for Space-Earth Environmental Research, Nagoya University, Furo, Chikusa, Nagoya, 464-8601, Japan}
\address[id=7]{National Institute of Information and Communications Technology, 4-2-1, Nukui-Kitamachi, Koganei 184-8795, Tokyo, Japan}
\address[id=8]{Astrophysics Group, Cavendish Laboratory, JJ Thomson Avenue, Cambridge CB3 0HE, UK}
\address[id=9]{European Southern Observatory, Alonso de C\'{o}rdova 3107, Vitacura 763-0355, Santiago, Chile}
\address[id=10]{European Southern Observatory (ESO), Karl-Schwarzschild-Strasse 2, 85748 Garching bei M\"unchen, Germany }
\address[id=11]{Hvar Observatory, Faculty of Geodesy, University of Zagreb, Ka\v{c}i\'{c}eva 26, HR-10000, Zagreb, Croatia}
\address[id=12]{Astronomical Institute, Academy of Sciences, Fri\v{c}ova 298, 251 65 Ond\v{r}ejov, Czech Republic}
\address[id=13]{Korea Astronomy and Space Science Institute, 776, Daedeokdae-ro, Yuseong-gu, Daejeon, 305-348, Republic of Korea}
\address[id=14]{School of Physics and Astronomy, University of Glasgow, Glasgow, G12 8QQ, Scotland、UK}
\address[id=15]{Space Sciences Laboratory, University of California, Berkeley, 7 Gauss Way, Berkeley, CA 94720, USA}
\address[id=16]{Institute of Theoretical Astrophysics, University of Oslo, Postboks 1029 Blindern, N-0315 Oslo, Norway}
\address[id=17]{Center For Solar-Terrestrial Research, New Jersey Institute of Technology, Newark, NJ 07102, USA}
\address[id=18]{Lockheed Martin Solar \& Astrophysics Lab, Org. A021S, Bldg. 252, 3251 Hanover Street, Palo Alto, CA 94304, USA}
\address[id=19]{Max-Planck-Institut for Sonnensystemforschung, Justus-von-Liebig-Weg 3, 37077 G\"{o}ttingen, Germany}
\address[id=20]{Astronomical Institute, St.Petersburg University, Universitetskii pr. 28, 198504 St.Petersburg, Russia}
\address[id=21]{Center for Space Plasma and Aeronomic Research, Univ. of Alabama Huntsville, Huntsville, AL 35899, USA}
\address[id=22]{National Astronomical Observatories, Chinese Academy of Sciences, A20 Datun Road, Chaoyang District, Beijing 100012,China}

%
\runningauthor{M. Shimojo {\it et al.}}
\runningtitle{ALMA Solar Interfereometry}


\begin{abstract}
Observations of the Sun at millimeter and submillimeter wavelengths offer a unique probe into the structure, dynamics, and heating of the chromosphere; the structure of sunspots; the formation and eruption of prominences and filaments; and energetic phenomena such as jets and flares. High-resolution observations of the Sun at millimeter and submillimeter wavelengths are challenging due to the intense, extended, low-contrast, and dynamic nature of emission from the quiet Sun, and the extremely intense and variable nature of emissions associated with energetic phenomena. The Atacama Large Millimeter/submillimeter Array (ALMA) was designed with solar observations in mind. The requirements for solar observations are significantly different from observations of sidereal sources and special measures are necessary to successfully carry out this type of observations. We describe the commissioning efforts that enable the use of two frequency bands, the 3 mm band (Band 3) and the 1.25 mm band (Band 6), for continuum interferometric-imaging observations of the Sun with ALMA. Examples of high-resolution synthesized images obtained using the newly commissioned modes during the solar commissioning campaign held in December 2015 are presented. Although only 30 of the eventual 66 ALMA antennas were used for the campaign, the solar images synthesized from the ALMA commissioning data reveal new features of the solar atmosphere that demonstrate the potential power of ALMA solar observations. The ongoing expansion of ALMA and solar-commissioning efforts will continue to enable new and unique solar observing capabilities.
\end{abstract}

%

\end{opening}


\section{Introduction}

The Atacama Large Millimeter/submillimeter Array (ALMA) is a powerful, general purpose radio telescope designed to address a broad program of forefront astrophysics at millimeter and submillimeter (mm/submm) wavelengths \citep{5136193,2010SPIE.7733E..17H}. Briefly, ALMA is an interferometric array that will ultimately be comprised of 66 antennas: $50 \times 12$ m antennas (the 12-m array); $12 \times 7$ m antennas (the 7-m array); and $4\times 12$ m ``total power" antennas (the TP array)\footnote{Atacama Compact Array (ACA: also known as the Morita Array) is a short-spacing imaging system consisting of the TP array and 7-m array \citep{2009PASJ...61....1I}.}. All antennas are capable of observing continuum and spectral line radiation at frequencies ranging from 35\ --\ 950 GHz (or wavelengths of 0.32\ --\ 8.6 mm). The 12-m array is reconfigurable, with the distance between two antennas (the antenna baseline) ranging from 15 m up to 16 km, thereby providing great flexibility in angular resolution and surface brightness sensitivity. The 7-m array is a compact fixed array with baselines ranging from 9 m to 50 m. The four TP antennas are used as single dishes to measure emission on the broadest angular scales. The 7 m antennas bridge the gap between the angular scales measured by the TP antennas and those measured by the 12-m array. ALMA is located on the Chajnantor plain of the Chilean Andes at an elevation of 5000 m, an exceptional site for mm/submm observations. 

ALMA science operations began with Cycle 0 in October 2011 with limited numbers of antennas and capabilities. Both the instrument and its capabilities have been expanding steadily since then and for ALMA observing Cycle 4, beginning in October 2016, at least $40\times 12$ m antennas, $10\times 7$ m antennas, and 3 TP antennas were available for scientific observations. The 12-m array supported nine antenna configurations and a total of seven frequency bands was available for scientific use on all antennas in Cycle 4. Additional technical details are available in the ALMA Cycle 4 Technical Handbook\footnote{http://almascience.org/documents-and-tools/cycle4/alma-technical-handbook}\citep{ALMAC4Tech}. A more complete description of solar observing modes supported by ALMA in Cycle 4 is given in Section 5.

Solar physics has been an important component of the ALMA science program since its inception. Continuum and spectral-line radiation from the Sun at mm/submm wavelengths offers a unique probe of chromospheric structure and dynamics; the structure and dynamics of sunspots; of the formation and eruption of prominences and filaments; and of energetic phenomena such as jets and flares \citep[see, \eg][]{2002AN....323..271B,2011SoPh..268..165K}. Particularly powerful are submm/mm observations carried out jointly with optical/IR and UV/EUV observations. A recent and comprehensive overview of solar science with ALMA is presented by \citet{2016SSRv..200....1W}. 

Although the antennas of ALMA were carefully designed and constructed for observing the Sun, it is not possible to observe the Sun using the same observing modes that are employed for other astronomical objects. Since the other components of ALMA are optimized to observe faint objects (\eg high-z galaxies), the mm/sub-mm wave radiation from the Sun is outside the normal operating parameter range. In addition, in the case of ALMA, the Sun is significantly bigger than the field of view (primary beam) of the ALMA 7 m and 12 m antennas, and the field of view is filled with complex solar structures that occupy a wide range of spatial frequencies. For these reasons, special measures must be taken to observe the Sun with ALMA. 

To open solar observing to researchers, the ALMA solar development team, the joint ALMA observatory (JAO), and ALMA regional centers (ARC) of East Asia, Europe, and North America have been developing and commissioning solar observing modes that exploit both single-dish total-power maps of the Sun and interferometric observations of the Sun.  Six solar commissioning campaigns were conducted from 2011\ --\ 2015, culminating with the release of ALMA Band 3 and Band 6 to the solar community for continuum observing in the Cycle 4 proposal cycle \citep[October 2016 \ --\ September 2017,][]{ALMAC4PropG}. 

In this article we present an overview of the observing modes available for interferometric solar observations with the Band 3 and Band 6 receivers \citep{2008SPIE.7020E..1BC,2004stt..conf..181E} of ALMA in Cycle 4. A companion article by \citet{White17} presents techniques developed for rapidly mapping the Sun  using the TP antennas. In Section 2, we explain the challenges of observing the Sun with ALMA and the steps taken to address them. In Section 3 we discuss calibration procedures developed for solar observing. In Section 4 we present some Scientific Verification data (SV data) of the Sun that were released by the JAO in January 2017. The SV data were obtained during the sixth ALMA solar commissioning campaign, held in December 2015. We conclude in Section 5 with a brief discussion of future solar capabilities with ALMA. 

\section{Solar Observing with ALMA}

In this section, the particular problems posed by observing the Sun with ALMA are outlined and their resolution is discussed. To include the Sun as part of ALMA's scientific program meant designing telescope hardware, electronics, and computing systems that could achieve high performance across an extraordinary range of spatial, spectral, temporal, and intensity scales without compromising the performance of any mechanical, electrical, or optical element along the signal path. A critical problem for observing the Sun with a precision telescope like ALMA is the potential thermal load on the antennas imposed by the optical and infrared (OIR) radiation from the Sun. The issue was considered carefully during the design phase of ALMA, and it was mitigated by ensuring that the dish panel surfaces are rough enough at OIR wavelengths to scatter the bulk of the OIR radiation out of the optical path \citep{2004SPIE.5489.1085U,2006PASP..118.1257M,5136193} while maximizing the antenna efficiency at mm/submm wavelengths (better than 25 $\mu$m rms surface accuracy). Therefore, we do not discuss the issue further in this article. 

\subsection{Reduction of the Solar Signal at mm/submm Wavelengths}

The Sun is an intense mm/submm source, orders of magnitude more intense than cosmic sources that ALMA is optimized to observe. The brightness temperature of the Quiet sun is 5000\ --\ 7000 K in the ALMA frequency, range and active-Sun phenomena can produce much higher brightness temperatures. ALMA receivers are designed for a maximum signal corresponding to an effective brightness of about 800 K at the receiver input, thereby limiting their dynamic range. Therefore, the solar signal must be attenuated or the receiver gain must be reduced to ensure that receivers remain linear, or nearly so. Two approaches to this problem were developed and tested during the commissioning phase: i) to attenuate the signal with a solar filter placed in the optical path in front of the receiver; or ii) to reduce the receiver gain to provide it with greater dynamic range. We discuss each approach in turn.     

\subsubsection{ALMA Solar Filters}

The initial solution adopted by ALMA to manage the input signal was the use of a solar filter (SF) that is mounted on the Amplitude Calibration Device (ACD) of each antenna \citep{ALMAC4Tech}. When placed in the optical path the solar filter is required to attenuate the signal by $4+2\lambda_{\rm mm}$ dB with a return loss of -25 dB (-20 dB for $\nu> 400$ GHz) and a cross polarization induced by the filter of -15 dB, or less.  There are several drawbacks to this solution \citep[see][]{6665775}: 

\begin{itemize}
\item The hot and ambient calibration loads on the ACD cannot be observed when the SF is in the optical path, making amplitude calibration difficult.
\item The SNR on calibrator sources is greatly reduced, not just by the attenuation introduced by the SF, but by the thermal noise that is added to the system temperature by the SF itself. 
\item The SF introduces frequency-dependent (complex) gain changes that may be time dependent and must be calibrated.
\item The SFs introduce significant wave-front errors into the illumination pattern on the antenna, resulting in distortions to the beam shape and increased sidelobes.
\item The Water Vapor Radiometers (WVRs) are blocked by the ACD for many bands when the SF is inserted into the optical path and phase corrections based on WVR measurements are therefore not possible in these bands.
\end{itemize}

Some of these difficulties have been overcome, \eg the complex gains of antennas outfitted with SFs were measured during the third solar observing campaign in 2013, and interferometric imaging with solar filters has been demonstrated. In fact, the SFs will likely be used for observations of solar flares at some future time. Nevertheless, the disadvantages to the use of solar filters are significant. They must be moved out of the beam when observing calibrators, thereby increasing operational overhead. Since they introduce frequency-dependent and possibly time-dependent gains, they must be measured for every filter and frequency setting. Other calibrations including pointing, focus, and beam-shape measurements need to have the filters in place. The reduced SNR makes such measurements more difficult and time consuming. 

\subsubsection{Receiver Gain Reduction}

While the use of solar filters has been demonstrated to work, their use introduces enough disadvantages to consider whether an alternative approach may be more attractive. \citet{6665775} pointed out that the ALMA Superconductor-Insulator-Superconductor (SIS) mixers could be de-tuned or de-biased to reduce the mixer gain. Since the dynamic range scales roughly inversely with gain, these settings can handle larger signal levels before saturating, potentially allowing solar observing without the use of the SFs, at least for non-flaring conditions on the Sun. This idea is illustrated in Figure 1, which shows the SIS current (left axis) and conversion gain (right axis) plotted against the SIS voltage bias for the ALMA Bands 3 and 6 receivers. The normal voltage bias tuning is on the first photon step below the gap where the gain conversion has a maximum. However, the mixer still operates at other voltage bias settings. These settings produce lower conversion gain, but since the dynamic range scales roughly inversely with gain, these settings can handle larger signal levels before saturating. In addition to the SIS bias voltage, the local oscillator (LO) power can be altered in order to further modify the receiver performance although that has not been explored in detail. 

\begin{figure}[t]
\centering
\includegraphics[scale=0.8]{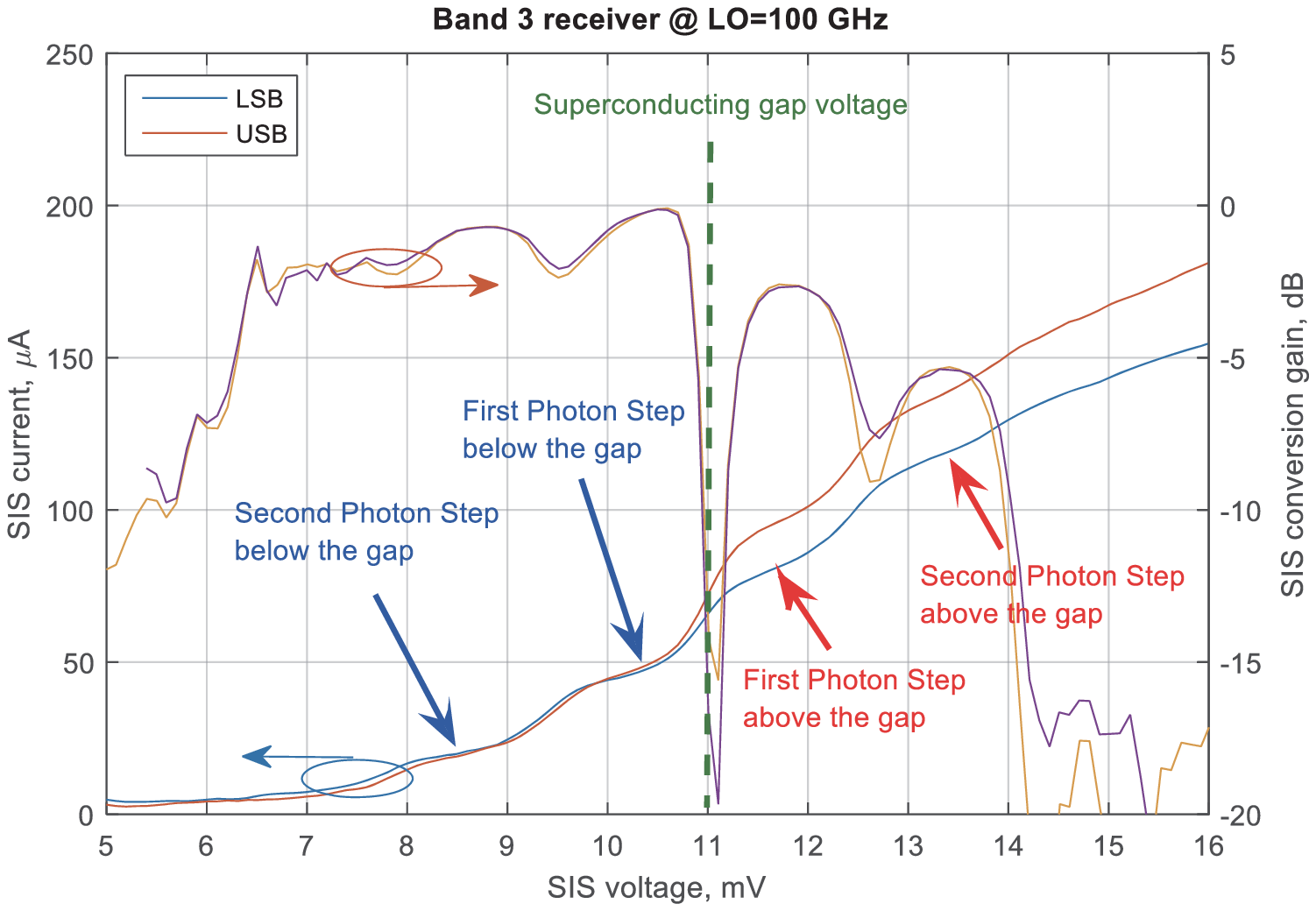}
\includegraphics[scale=0.8]{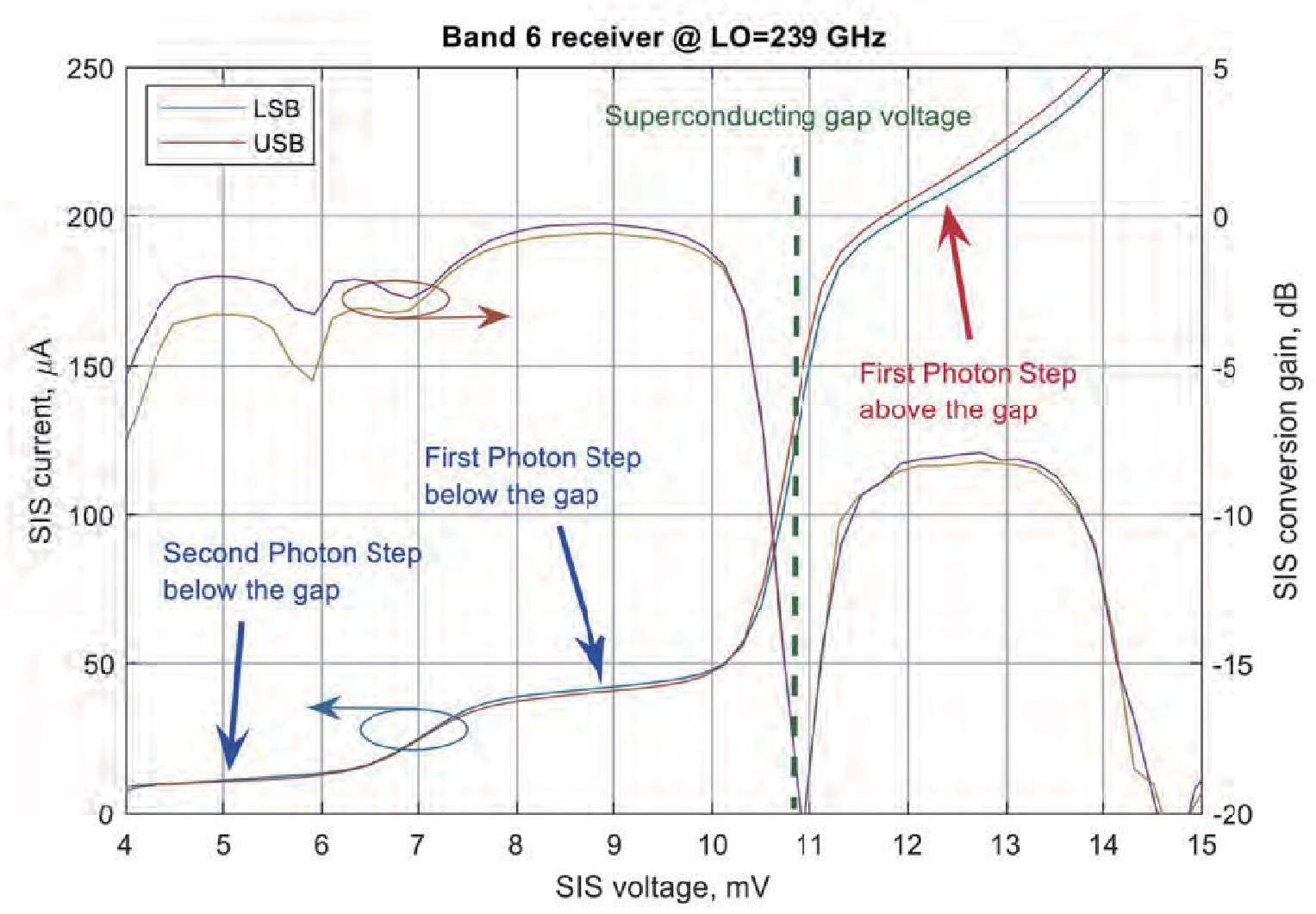}
\caption{SIS current and conversion gain as a function of voltage setting for the ALMA Band 3 (left) and Band 6 (right) receivers. The arrowed ellipses indicate the relevant ordinate: left for the SIS current and right for the conversion gain. \label{fig:fig1}}
\end{figure}

Tests conducted in 2014 showed that for Band 3 the second photon step below the gap has the flattest gain response as a function of SIS bias voltage as well as better linearity and sensitivity than the first step above the gap. This second photon step below the gap is suitable for ``quiet" Sun observations and is referred to as ``Mixer-Detuned mode 1", or MD1. For ``active" Sun observations, further gain reduction is achieved by tuning to the second photon step above the gap, referred to as MD2. For Band 6 receivers it was found that the second photon step below the gap did not always provide a flat and stable gain response (at least with nominal LO power). Moreover, the receiver gain compression is moderate on the quiet Sun even at nominal receiver settings (first step below the gap). Therefore, no change from nominal settings is recommended for ``quiet" Sun observing (MD1). For ``active" Sun observations, tuning to the first photon step above the gap is recommended (MD2). It is seen that for both Band 3 and Band 6 receivers the MD2 mode provides reduced gain and better linearity for quiet Sun inputs. However, the improved dynamic range comes at the cost of higher system temperature due to increased receiver noise. This is not a problem when pointing at the Sun, for which the antenna temperature is significantly larger than the system temperature (see Section 3.1).

\begin{table}[ht]
\caption{SIS Mixer Settings for ALMA Cyle 4 solar observations}
\begin{tabular}{lcccc}
\hline
{\bf Band 3}  & 2nd Step Below Gap & 1st Step Below Gap & 1st Step Above Gap & 2nd Step Above Gap \\
       &  (MD1)  &  (nominal)  &    &  (MD2)  \\
\hline
Receiver noise [K] & $\approx$50 & $<$41 & $\approx$200 & $\approx$800 \\
Estimated compression & $\approx$10\ \%  & $\approx$35\ \% & $\approx$15\ \%  & (a few \%) \\
(Quiet Sun input) &&&& \\
\hline
{\bf Band 6}  & 2nd Step Below Gap & 1st Step Below Gap & 1st Step Above Gap &  - \\
       &   & (nominal/MD1)   &  (MD2)   &   \\
\hline
Receiver noise [K] & $\approx$200 & $<$83 & $\approx$1000 & - \\
Estimated compression & $<$5\ \% & $\approx$10\ \% & (a few \%) &  - \\
(Quiet Sun input) &&&& \\
\hline
\end{tabular}
\end{table}

The analyses by \citet{Yagoubov16,Iwai15,Iwai16} report that the gain compression, an indicator of non-linear response of the receiver, $\approx$10\ \% at quiet sun and $\approx$15\ \% at active regions for the MD1 mode in both the Band 3 and Band 6 receivers. Considering the specification of the receivers, the decreasing of the sensitivity, and the brightness temperature range of the Sun, the receivers with the MD2 mode are believed to respond linearly. However, we cannot exclude the possibility that there is small amount of gain compression, which could possibly be a few \%. Taking account of the current precision of solar visibilities obtained with ALMA, the influence of the nonlinear response with the MD2 mode is sufficiently small as to be neglected. On the other hand, if the MD1 mode is used to observe active regions and flares, the mildly nonlinear response will reduce the accuracy of measured brightness temperatures. If the MD2 mode is used to observe flares it, too, may suffer significant gain compression. 

In assessing the two approaches to managing solar signals, it was concluded that the use of MD modes is preferable over the use of SFs for the non-flaring Sun because of the greater simplicity of their implementation and the associated calibration procedures, as we discuss further below. That said, it is likely that the use of SFs will be necessary to observe solar flares at mm/submm wavelengths. 

\subsection{Managing Signal Power Prior to Digitization}

Calibration of the ALMA antenna gains is discussed in greater detail in Section 3. Briefly, a calibrator source with known properties is observed by the array and the complex gains are deduced. The phase solutions are then transferred to the source data. For reasons discussed in Section 3, it is both possible and desirable to observe both the Sun and calibrator sources in a fixed MD mode. While both calibrators and the Sun can be observed in an MD mode, the power entering the system when pointing to cold sky when observing a calibrator and the power entering the system when observing the Sun are vastly different. ALMA employs two stages of heterodyne frequency conversion to shift the observed (sky) frequency down to a frequency range where system electronics can be used to digitize the analog signals and then correlate them. First, the signals at the observed radio frequency on the sky are mixed with a reference frequency (local oscillator (LO)) to an intermediate frequency (IF). The resulting IF frequency bands lie above and below the LO frequency (upper sideband and lower sideband). These are further subdivided and down-converted with a second LO in the IF Processor to four basebands that lie within the 2\ --\ 4 GHz band. For continuum observations in ALMA Bands 3 and 6, a total of four 2 GHz bands are processed. These are referred to as spectral windows (SPWs). The continuum spectral windows observed by ALMA in Band 3 and Band 6 for Cycle 4 are detailed in Table 2.
\medskip

\begin{table}[ht]
\caption{Continuum frequencies for ALMA Cycle 4 solar observations}
\begin{tabular}{lcccc}
\hline
       & SPW 1 & SPW 2 & SPW 3 & SPW 4 \\
\hline
Band 3 &  92\ --\ 94 GHz & 94\ --\ 96 GHz & 104\ --\ 106 GHz & 106\ --\ 108 GHz\\
Band 6  & 229\ --\ 231 GHz & 231\ --\ 233 GHz & 245\ --\ 247 GHz & 247\ --\ 249 GHz\\
\hline
\end{tabular}
\end{table}

The baseband signals are then digitized and correlated. The analog-to-digital converters (ADCs) are sensitive to input power; an important consideration given the difference in input power when observing calibrators and the Sun because the difference exceeds the dynamic range of the ADCs by a large margin. To adjust the input levels to the ADC to optimum values it is necessary to adjust signal power levels through the use of two stepped attenuators under digital control. One stepped attenuator is in the IF Switch, which controls which receiver signal enters the IF Processor; the other stepped attenuator is in the IF Processor itself \citep{ALMAC4Tech}. 

The solar development team carried out extensive test observations in October and November 2014 to determine the appropriate attenuator values. The stepped attenuators were set to values that optimized ADC signal input levels when observing the Sun. However, when the attenuation levels configured in the IF Switch and IF Processor are optimized for the Sun, they are non-optimum for calibrator sources.  It is necessary to reduce the attenuation levels relative to the solar values when observing phase and bandpass calibrators. The recommended input level to the ADCs is 3.8 dBm. By adjusting IF Switch and IF Processor attenuation levels for calibrator observations relative to those used for observations of the Sun the input levels into the ADCs for observations of both the Sun and calibrators are near the recommended value (Table 3).

\begin{table}[ht]
\caption{Differences of the attenuation levels for the calibrators from those for the Sun}
\begin{tabular}{lccccc}
\hline
&& \multicolumn{2}{c}{Diff'l Attenuation} & \multicolumn{2}{c}{Input Level to ADCs} \\
Receiver & MD mode & IF Switch& IF Proc & Sun & Calibrator (sky) \\
\hline
Band 3 & MD1 & -8 dB & -10 dB & $\approx\!3.5$ dBm & $\approx\!3.5$ dBm\\ 
 & MD2 & -8 dB & 0 dB & $\approx\!3$ dBm & $\approx\!4$ dBm\\ 
Band 6 & MD1 & -10 dB & -10 dB & $\approx\!3.5$ dBm & $\approx\!2.5$ dBm\\
 & MD2 & -8 dB & 0 dB & $\approx\!4$ dBm & $\approx\!4.5$ dBm\\
\hline
\end{tabular}
\end{table}

\section{Solar Data Calibration}

An interferometric measurement of the source at a particular time, frequency, and polarization by a pair of antennas is referred to as a ``visibility". It is a complex quantity characterized by an amplitude and a phase and may be thought of as a single spatial Fourier component of the brightness distribution of the source. The measurement is made in the aperture plane : the \uv-plane. The objective is to sample the \uv-plane with sufficient density to recover the brightness distribution of source through Fourier inversion of the visibilities coupled with deconvolution techniques. Two key calibrations of ALMA visibility data are to measure the (time variable) complex gain of each antenna (amplitude and phase calibration) and to place the measurements on an absolute flux scale (flux calibration). Many additional calibrations are routinely required: antenna baselines, delay, frequency bandpass, polarization, etc. Those possibly affected by solar observing are touched on below.

\subsection{Gain Calibration}

Normally, amplitude and phase calibration of the antenna gains are performed by observing strong mm/ submm sources with accurately known positions, structure, and flux densities. Most flux calibrators are strong quasars or planets while phase calibrators are usually quasars that are point-like to the antenna array. By observing a phase-calibrator source every few minutes, the complex antenna gain (amplitude and phase) is deduced as a function of time. The gain solutions are then interpolated to the times at which the solar source is being observed and applied to the source data. The overall flux scale is determined by scaling the visibilities to Kelvin using measurements of the System Equivalent Flux Density (see Section 3.2) and further referencing the scaled visibility measurements to those of the flux calibrator \citep[see Section 10.5 of][]{ALMAC4Tech}. However, the calibration of solar observations differs in key respects from those of faint, non-solar sources as we now discuss.

\begin{figure}[t]
\centering
\includegraphics[scale=.6]{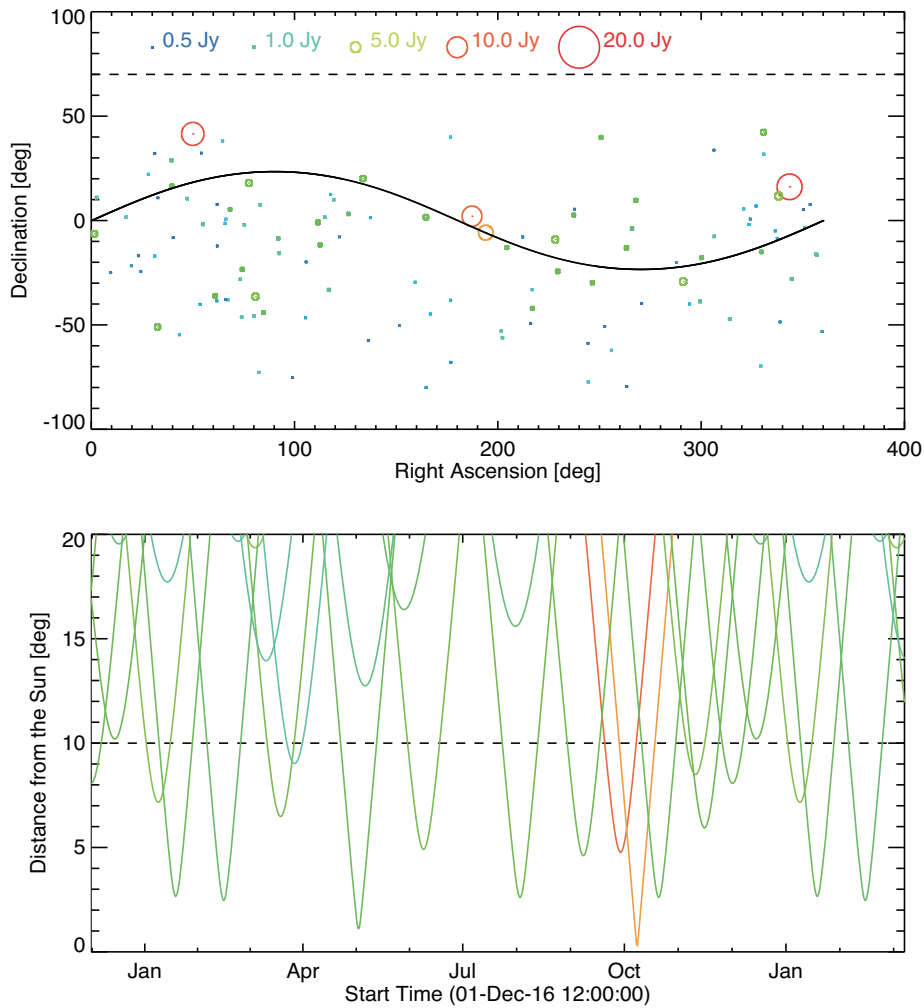}
\caption{Upper panel: The distribution of the quasars brighter than 0.5 Jy in Band 6. The color and size of the circle indicates the flux of a quasar. The black line indicates the track of the Sun. Lower panel: The separation angle between the Sun and possible calibrator sources ($>\!1$ Jy). The color indicates the flux of a quasar (same as that used in the upper panel).\label{fig:fig2}}
\end{figure}

When the SIS mixers are de-tuned to an MD mode the dynamic range of the receivers can accommodate the strong signal input from the Sun in a (nearly) linear fashion. Adopting the so-called MD mode for solar observing comes with two penalties: first, by tuning away from the nominal bias voltage in the SIS mixer, the MD mode introduces an unknown, but stable, gain change to the signal. This can either be measured for each antenna, frequency, and polarization or it can be ignored by observing both the source and the calibrator using the MD mode, in which case the gain change cancels out. The latter approach has been taken. Second, the use of MD modes results in an increase in receiver noise and a corresponding reduction in sensitivity. While the MD1 mode results in only a modest increase in the receiver noise ($\approx 20\ \%$ in band 3) the use of MD2 mode results in a much more significant increase of the receiver noise: characterized in terms of the receiver temperature, it increases from $\approx 50$ K to $\approx 1000$ K. This is not a problem as long as sufficiently strong calibrator sources are available that can overcome the reduced sensitivity. In practice, strong calibrators become increasingly sparse, especially at higher frequencies and so care must be taken in identifying a suitable calibrator source when using the MD2 mode. Using the ALMA Calibrator Source Catalogue, Figure 2  shows the distribution of possible calibrators that can be observed with the MD2 mode in Band 6 as a function of their flux density and position relative to the Sun (solid black line). There is a period in early July when there is no suitable calibrator within $20^\circ$ of the Sun, which leads to degraded transfer of phases during calibration (see Section 4). The situation is similar for Band 3; hence, observations of the Sun with the frequency bands higher than Band 3 are not recommended in early July.

As described in Section 2, steps have been taken to ensure that input signals remain nearly linear and within power limits to ensure optimum system performance when observing the Sun and calibrator sources. However, in addition to maintaining signal levels one must ensure that the signal phase is maintained. Phase differences between calibrator and solar-source scans are avoided by using the same MD mode to observe both. However, an additional concern is whether the stepped IF Switch and IF Proc attenuators themselves introduce unacceptable system temperature changes and/or differential phase variation between the Sun and calibrator settings, thereby corrupting phase calibration referenced against suitable calibrator sources. The variation in system temperature caused by the stepped attenuators is negligibly small, so it is not necessary to correct for their influence on flux calibration. On the other hand, the stepped attenuators do introduce significant phase shifts, depending on the difference in attenuation introduced for solar and calibrator scans. If the values of the phase shifts in all of the antennas are identical, however, the phase shift will be differenced out and the transfer of phase calibration from a calibrator to the solar source can proceed without the added complexity of measuring and applying differential phase corrections to account for phase errors introduced by the IF Switch and IF Processor attenuators. To check this, the bright quasar 3C279 was observed during the commissioning campaign in December 2015 while systematically changing IF Switch and IF Proc attenuator states on all antennas. Figure 3 shows an example of the differential phase variations caused by changing the attenuation levels. The channel-averaged value of the phase variation in a spectral window is very close to 0, and its standard deviation across the spectral window is 0.3 degrees for the attenuator in the IF Switch, and 0.6 degrees in the IF Processor. Moreover, there is no significant change of the phase variation during the campaign. These results indicates that the characteristics of the stepped attenuators are uniform and stable, and the phase variation caused in one antenna is almost canceled out by that in the other antenna. Therefore, there is no need to carry out additional calibration for the phase variation caused by changing the attenuation levels.

\begin{figure}[t]
\centering
\includegraphics[scale=1.3]{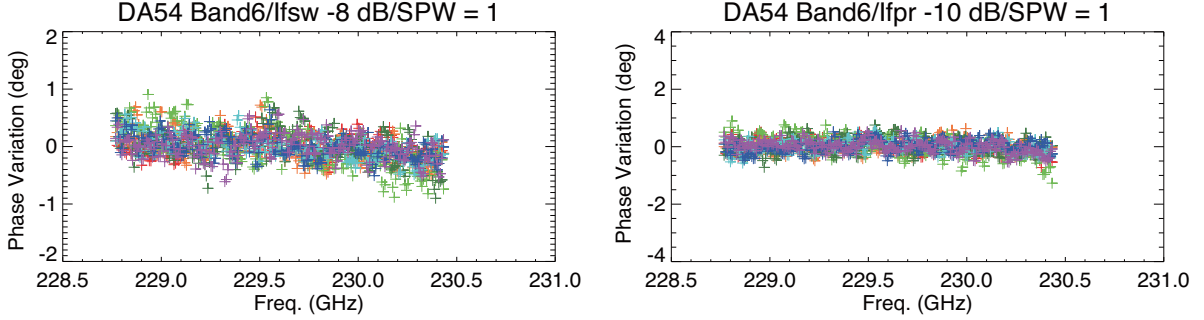}
\caption{The (differential) phase variation in a Band 6 spectral window. Left: The case of changing the attenuator in IF Switch -8 dB from the solar setting. Right: The case of changing the attenuator in the IF Processor by -10 dB. Colors indicate the observing day; red: 14, orange: 15, green: 16, dark green: 17, blue:18, purple:20 December 2015. \label{fig:fig3}}
\end{figure}

As we shall see below, amplitude and flux calibration referenced to standard source such as strong quasars or planets do not apply to solar data. The reason is that, in contrast to the vast majority of sidereal sources, the ``antenna temperature" $T_{\rm ant}$, which indicates the input power from an observing target (Sun) in equivalent temperature scale, is significantly larger than the ``system temperature" $T_{\rm sys}$ that indicates the system noise due to the receiver, other electronics, and spurious signals. In addition, the Sun is obviously not point-like; it fills the field of view of both the 7 m and 12 m antennas and their sidelobes. To properly calibrate visibility amplitudes and place them on a common flux scale it is necessary to measure both $T_{\rm sys}$ and $T_{\rm ant}$. 

\subsection{Flux Calibration}

When an astronomical object is normally observed with ALMA, the output from the correlator is a normalized cross-correlation coefficient $\rho_{\rm mn}$ for a pair of antennas $m$ and $n$, is written as 

\begin{equation}
\rho_{\rm mn} = {\sqrt{T_{\rm corr_m}T_{\rm corr_n}}
\over 
{\sqrt{(T_{\rm ant_m}+T_{\rm sys_m})(T_{\rm ant_n}+T_{\rm sys_n})} }  }
\end{equation}

\noindent where $T_{\rm ant}$ is the antenna temperature, $T_{\rm sys}$ is the system temperature, and $T_{\rm corr}$ is the temperature of the correlated component of $T_{\rm ant}+T_{\rm sys}$. The relation between antenna temperature [in units of K] and flux density $S$ [units W Hz$^{-1}$ m$^{-2}$] is 

\begin{equation}
T_{ant} = { {SA_{\rm e}} \over {2k} }
\end{equation}

\noindent where $k$ is the Boltzmann constant and $A_e$ is the effective antenna collecting area [$m^2$]. The relation is also valid for the correlated component. From Eqs. (1) and (2), the amplitude of a visibility measurement is 

\begin{equation}
S_{\rm corr_{mn}} = 2k {  
{\sqrt{(T_{\rm ant_m}+T_{\rm sys_m}) (T_{\rm ant_n}+T_{\rm sys_n})}}
\over 
{\sqrt{A_{\rm e_m} A_{\rm e_n}}}
} \rho_{\rm mn}
\end{equation}

\noindent A System Equivalent Flux Density (SEFD) is defined as 

\begin{equation}
SEFD = 2k { {T_{\rm sys}} \over {A_{\rm e}} }
\end{equation}

\noindent Then, the amplitude of a visibility is written as

\begin{equation}
S_{\rm corr_{mn}} = \rho_{\rm mn}\sqrt{SEFD_{\rm m}\ SEFD_{\rm n}}\sqrt{(1+q_{\rm m})(1+q_{\rm n})}
\end{equation}

\noindent where $q=T_{\rm ant}/T_{\rm sys}$. The antenna temperature of most celestial sources is much smaller than the system temperature, $T_{\rm ant} \ll  T_{\rm sys}$, and $q = 0$. This is the case for calibrator sources which only need measurements of $T_{\rm sys}$ to scale the visibilities. In contrast, when observing the Sun $T_{\rm ant}>T_{\rm sys}$, and it is therefore necessary to measure both $T_{\rm sys}$ and $T_{\rm ant}$ in order to correctly scale the visibility measurements. The procedure for measuring $T_{\rm ant}$ and $T_{\rm sys}$ is described in detail by \citet{White17} in the context of single dish observations of the Sun. Briefly, the antenna temperature is measured using the ACD on which ``hot load" and ``cold load" reference inputs are available. The following measurements are performed before each source scan:

\begin{itemize}
\item a cold-load observation $P_{\rm cold}$ (also known as the ambient load), in which an absorber at the temperature of the thermally controlled receiver cabin (nominally $15\ --\ 18^\circ$ C) fills the beam path;
\item a hot-load observation $P_{\rm hot}$, in which an absorber heated to about $85^\circ$ C fills the beam path
\item a sky observation $P_{\rm sky}$, offset from the Sun (typically by two degrees) and at the same elevation. The attenuation levels of the attenuators in IF chain are the same as that for the measurement of $P_{\rm cold}$ and $P_{\rm hot}$.
\item an off observation $P_{\rm off}$, which is the same as the $P_{\rm sky}$, except the attenuation levels are set to the values optimized for the Sun
\item a Sun observation $P_{\rm sun}$, which is at the attenuation levels of the target (Sun)
\item a zero level measurement $P_{\rm zero}$, which reports the levels in the detectors when no power is being supplied.
\end{itemize}

The autocorrelation data output from the correlator cannot be used for these measurements because the correlator has insufficient dynamic range to measure $P_{\rm off}$. Instead, the measurements rely on the total-power data obtained by the baseband square-law detectors. The antenna temperature of the science target on the Sun is given by:

\begin{equation}
T_{\rm ant}=\frac{P_{\rm sky}-P_{\rm zero}}{P_{\rm off}-P_{\rm zero}} \frac{P_{\rm sun}-P_{\rm off}}{P_{\rm hot}-P_{\rm cold}} (T_{\rm hot}-T_{\rm cold})
\end{equation}
\smallskip

\noindent and the system temperature is given from the online measurements \citep[see Section 10.4,][]{ALMAC4Tech}. Additional details regarding flux calibration are provided in \citet{White17}.

\begin{figure}[t]
\centering
\includegraphics[scale=.7]{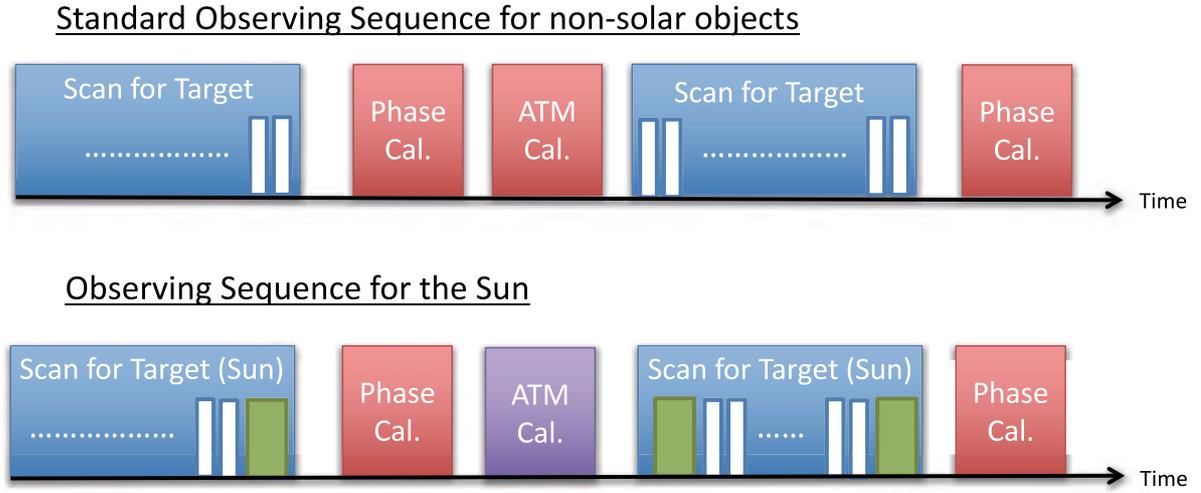}
\caption{A cartoon of the observing sequences around a scan of a target. A box indicates the period of a scan. A scan is constructed from multiple subscans as shown in the ``Scan for Target". Except the scan for target, subscans are omitted in the figure. A red box indicates a scan for the phase calibrator; a blue box is the scan for the scientific target, and a purple box indicates a scan for atmospheric calibration near the target. White narrow boxes indicate subscans of observing the target, and green boxes indicate subscans of observing sky near the Sun with the attenuating levels for observing the Sun.}
\end{figure}

\begin{figure}[h]
\centering
\includegraphics[scale=.6]{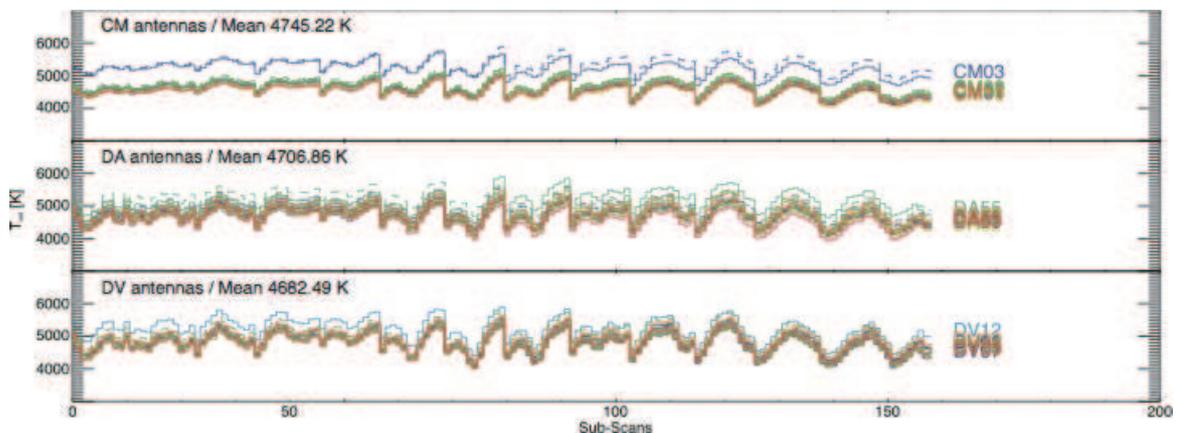}
\caption{The temporal variation of the antenna temperature as a function of subscan number during the 149-point mosaic observation of a sunspot with Band 6 using the MD2 observing mode. Upper panel: CM antennas (East Asia 7-m antennas), Middle panel: DA antennas (European 12-m antennas), Lower panel: DV antennas (North American 12-m antennas). Colors indicate the antennas. The solid and dashed lines indicate the polarization X (solid) and Y (dashed) .}
\end{figure}

To derive $T_{\rm ant}$ in practice requires modifying the standard ALMA observing sequence. There are three major differences, as shown schematically in Figure 4. The first is that subscans are needed for observing the sky near the Sun at the start and end of a science-target scan, the reason being that the $P_{\rm off}$ measurement has to be carried out with the attenuator levels set for observing the Sun. Hence, $P_{\rm off}$ is measured by the first and last subscans within the science target scan. The duration of the subscans used for measuring $P_{\rm off}$ is currently set to a few seconds. The second difference from standard procedures is that an atmospheric calibration is not carried out for each calibrator scan because it introduces too long a delay (many minutes)  between source scans, to the possible detriment of a given observer's scientific objectives. Instead, the system temperature derived from the atmospheric calibration near the Sun is applied to phase-calibrator data. The third modification is that the measurement of $P_{\rm zero}$ is carried out at the beginning of a solar observation. The value of $P_{\rm zero}$ is found to be very stable for a given antenna and frequency band, so we do not need to carry out the measurement frequently.  Since the subscan duration of the SV data is less than 30 seconds, a measurement of $T_{\rm ant}$ and the amplitude calibration of the visibilities are done for every subscan within a science target scan. Considering dynamic solar phenomena with short temporal scales (\eg flares), the short integration time for calculating $T_{\rm ant}$  is suitable for science although significant computer resources are needed for the amplitude calibration. Figure 5 shows an example of the time variation of $T_{\rm ant}$ for a mosaic observation, where an image is constructed from a pattern of discrete antenna pointings (see Section 4). It is clear that $T_{\rm ant}$ varies as ALMA points to different locations on the Sun.

\subsection{Bandpass Calibration}

Continuum observations are performed in four spectral windows. In fact, the observations in each SPW are coarsely channelized and corrected for the variation in phase and amplitude across the frequency band, a process that is referred to as bandpass calibration. Following bandpass calibration the channels may be summed and imaged as continuum emission. Bandpass calibration is carried out in the usual manner even when solar MD observing modes are used: \ie a strong calibrator is observed in an MD mode with the attenuator levels optimized for the Sun and the bandpass solution is obtained. The bandpass shape and stability were checked for the MD modes and attenuator states. It was found that the perturbations to bandpass amplitudes and phases were small. For the IF-switch and IF-processor-attenuator settings adopted for observations with an MD mode, it was found that the RMS difference between bandpass phases for an attenuator state and the nominal attenuator state was generally a fraction of a degree for both the Band 3 and Band 6 receivers, the maximum being 1.2 degrees. Similarly, the normalized amplitude difference was typically a fraction of 1\ \%. We conclude that no explicit correction is needed to normal bandpass calibration as a result of using MD modes or different attenuator states when observing calibrator sources and the Sun.

\subsection{Additional Considerations}

The primary beams of the main 12 m ALMA antennas are small compared with the Sun ($\approx 58$\arcsec\ at band 3, $\approx 24$\arcsec\ at band 6), and solar structures have various spatial scales. Therefore, to synthesize the solar brightness distribution visibility measurements should be distributed uniformly with spatial frequency in the aperture plane (the \uv-plane). The \uv-coverage can be improved by employing the Earth rotation synthesis technique, but this is only scientifically useful for slowly-varying, stationary structures, while many solar structures are dynamic in nature and vary on short time scales ($<$ one minute).
 
In Cycle 4, $40\times 12$ m antennas and $10\times 7$ m antennas were available for solar observing. The distribution of the 12 m antennas on the Chajnantor Plateau (array configuration) varies throughout the cycle from compact configurations to high-resolution long-baseline configurations. The proposal guide lists the configurations available for observing extended sources \citep[see Table A-2 of][]{ALMAC4PropG}. The table reveals that multiple configurations of the 12-m array are needed to observe extended sources in configurations larger than C40-4, as more extended configurations undersample the Sun's brightness distribution. Different configurations of the 12-m array cannot be realized at the same time. Therefore, solar observations must be carried out with the compact-array configurations. In Cycle 4, only the three most compact antenna configurations are available for solar observing: C40-1, C40-2, and C40-3. 

\begin{figure}[t]
\centering
\includegraphics[scale=0.6]{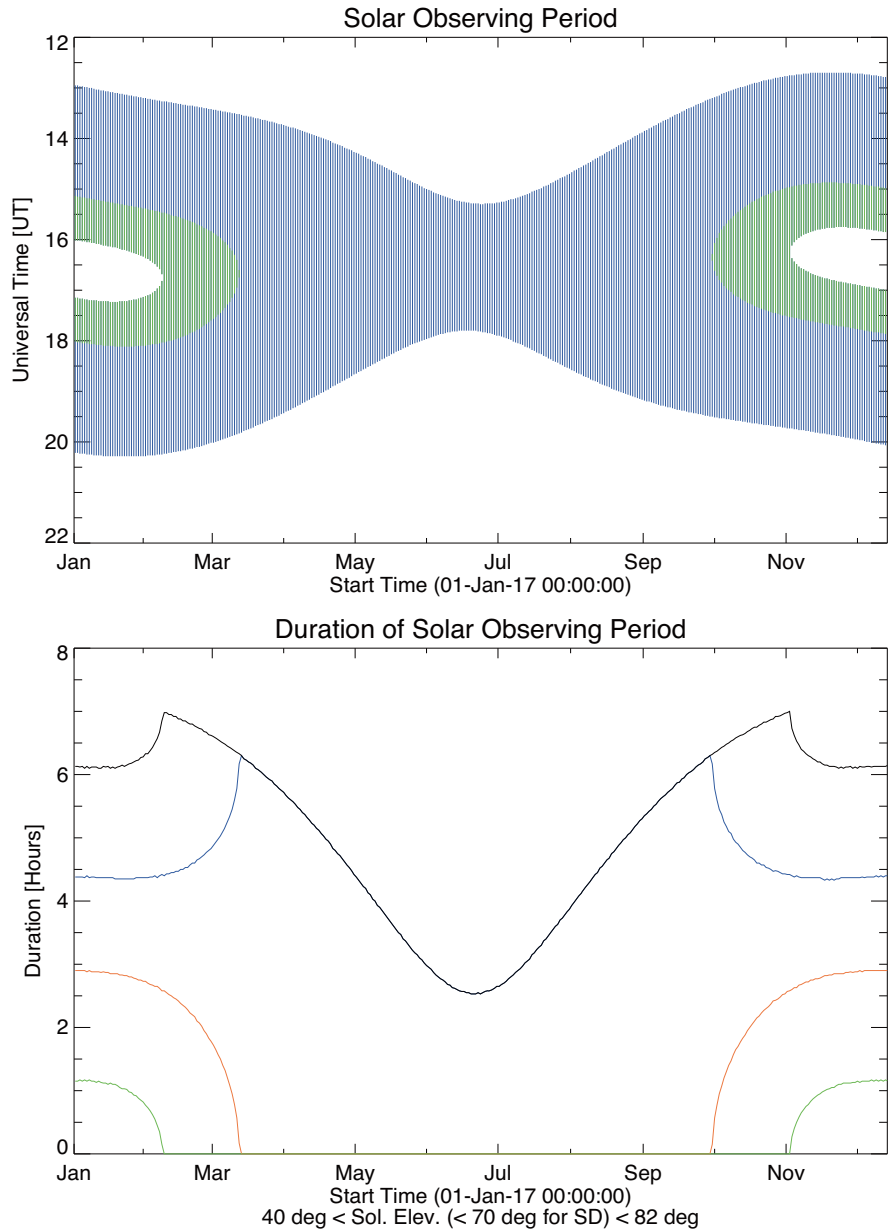}
\caption{The temporal range of solar observations. Upper panel: The blue lines indicate the possible temporal ranges of solar observations with both heterogeneous and Total-Power (TP) arrays. The green regions show the time range that we can use only heterogeneous array. Lower panel: Black line indicates the total duration of the solar observing with heterogeneous array in a day. Blue line shows the total duration of the solar observing with TP array in a day. Orange and green lines show the lost time caused by high elevation of the Sun (Orange: $>\!70^\circ$, Green: $>\!82^\circ$).}
\end{figure}

A second reason that solar observations are restricted to compact array configurations is that higher angular resolution requires longer antenna baselines, and longer baselines are susceptible to phase fluctuations caused by precipitable water vapor (PWV) in the atmosphere overlying the array. For non-solar observing, it is possible to estimate the amount of precipitable water vapor along the line of sight of each antenna using Water Vapor Radiometers \citep[WVR:][]{ALMAC4Tech}. Such measurements are essential for the phase calibration of long baselines, especially for higher frequencies. However, the WVR system is not available for solar observations because the WVRs saturate when the antennas point at the Sun. The issue of phase fluctuations on longer baselines and/or at higher observing frequencies will need to be confronted as new solar observing capabilities are made available -- for example, the use of band 7 (275\ --\ 373 GHz) and band 9 (602\ --\ 720 GHz).

Although solar observing is confined to compact array configurations, the spatial-frequency coverage of the \uv-plane from the 12 m antennas alone is still not adequate for synthesizing solar images. It is essential to observe the Sun with the 7-m array and 12-m array simultaneously. For non-solar observations, the 7-m array is operated with the ACA correlator \citep{2012PASJ...64...29K}. Simultaneous observations with the 7-m array and the 12-m array are not performed in general. However, since solar imaging requires the short baseline coverage provided by the 7-m array together with the longer baselines provided by the 12-m array both the 7m and 12m antennas are connected to the 64-input baseline correlator \citep{2007A&A...462..801E}, and the ACA correlator is not used for solar interferometric observations. In the other words, solar observations with ALMA are carried out with a heterogeneous array.

To synthesize solar images calibrated to absolute brightness temperatures, the data that are obtained from the heterogeneous array are still not complete because angular scales greater than those measured by the shortest antenna baselines are not available. These are measured by the TP antennas using fast-scan mapping techniques described by \citet{White17}. In Cycle 4, a solar observation with TP antennas are carried out with a solar interferometric observation simultaneously to enable the two types of measurement to be combined as appropriate. When the elevation of the Sun is higher than 70$^\circ$, we cannot observe the Sun with the fast-scanning mode of the TP array. On the other hand, to avoid shadowing, solar observations with the fixed 7-m array cannot be performed when the elevation of the Sun is lower than 40$^\circ$. Moreover, the heterogeneous array also cannot observe the Sun when the elevation is higher than 82$^\circ$. Considering these elevation limitations the temporal range for solar observations in a day is limited, as shown in Figure 6. 

\section{ALMA Solar Imaging Examples}

The solar commissioning campaign for verifying the Cycle 4 solar observing modes described above was held from 14\ --\ 21 December 2015. The specific modes and capabilities offered in Cycle 4, and verified during the campaign, are as follows:

\begin{itemize}
\item Band 3 and Band 6 continuum observations of the Sun will be supported
\item Solar observing will only be offered for the most compact array configurations
\item Both 7 m and 12 m antennas will be correlated by the 64-input baseline correlator
\item Both single pointing and mosaic (up to 150 pointings) interferometric observations of target sources will be supported
\item Observations with the interferometer will be supported by fast-scanning total power (TP) maps of the full disk of the Sun 
\end{itemize}

A number of solar targets was observed: active regions, quiet sun, solar limb, and a prominence above the limb. Only $\approx$30 antennas, including $9\times 7$ m antennas were typically available for the campaign. Therefore, the quality of the solar images presented in this article is not as good as those expected in Cycle 4 because of the larger number of antennas available in Cycle 4. Most of the data obtained from the December 2015 campaign were released by JAO as Scientific Verification (SV) data on 18 January 2017. The solar SV data can be downloaded from the ALMA Science Portal web site hosted by each ARC\footnotemark.

\subsection{Data and Image Synthesis}

In order to introduce solar images synthesized from ALMA observations, we use the SV data listed in Table 4. The observing period given in the table includes all calibrations required to execute the observation; \eg the bandpass and flux calibrations before the scientific scans. All of the examples given used the MD2 observing mode and the imaging employed the mosaic technique, in which a grid of discrete antenna pointings is used to image a much larger field of view than is available with a single pointing. For the examples presented here, the maximum number of mosaic pointings currently supported by the instrument were used: 149 pointings. The ICRS reference coordinates refer to the RA/Dec coordinates of the center of field of view at the reference time. The integration time for each MOSAIC pointing is 6.048 seconds, and the angular separation of points is 11.2\arcsec\ for Band 6, and 24.1\arcsec\ for Band 3; \ie Nyquist sampling in each Band.

\begin{table}[ht]
\caption{Science Verification Data Used}
\begin{tabular}{ccccccc}
\hline
Data Set & Execution Block ID & Frequency&Target & Observing& Reference & ICRS Reference\\
&&Band&&Period&Time&Coordinate\\
\hline
1&\textsf{uid://A002/Xade68e/X180}&3&AR 12470&18:01\ --\ 18:48UT&18:32:41UT&17$^{\rm h}$35$^{\rm m}$32.218$^{\rm s}$  \\
 & & && 16 Dec 2015 & &-23$^{\rm d}$16\arcmin23.843\arcsec \\
2&\textsf{uid://A002/Xae00c5/X2a8d}&6&AR 12470&19:15\ --\ 20:08UT&19:49:00UT&17$^{\rm h}$44$^{\rm m}$10.112$^{\rm s}$ \\
 & & && 18 Dec 2015 & & -23$^{\rm d}$19\arcmin30.632\arcsec \\
3&\textsf{uid://A002/Xae17cd/X367a}&6&South Pole&13:31\ --\ 14:32UT&14:09:37UT&17$^{\rm h}$51$^{\rm m}$46.086$^{\rm s}$ \\
 & & && 20 Dec 2015 & &-23$^{\rm d}$41\arcmin33.229\arcsec \\
\hline
\end{tabular}
\end{table}

\footnotetext{The URLs of the ``Scientific Verification Data" in each ARC web site are as follows:\\
EA-ARC: https://almascience.nao.ac.jp/alma-data/science-verification \\
EU-ARC: https://almascience.eso.org/alma-data/science-verification \\
NA-ARC: https://almascience.nrao.edu/alma-data/science-verification}

The Common Astronomy Software Applications (CASA) package \citep{2007ASPC..376..127M}, which is the standard reduction/imaging/analysis software for ALMA data, was used to calibrate and image the SV data. CASA can deal with data obtained with a heterogeneous array represented by the use of both 12m and 7m antennas. Hence, ALMA standard calibrating method is used for solar data, except for the amplitude calibration steps described in Section 3.2. When we use the {\sf clean} task of CASA for synthesizing a solar image, the {\sf mosaic} option for the {\sf imagemode} parameter has to be used even for the data of single-pointing observations, to deal with the heterogeneous-array nature of the data. For mosaic observations, the coordinate of each pointing has to be re-calculated relative to the center of the FOV using ALMA pointing data. This is necessary because the heliocentric coordinate frame is moving relative to the RA/Dec coordinate frame during an observation. The reference time in Table 4 indicates the time used to define the reference position of the Sun. To improve image quality, we include the data from all four SPWs for synthesizing one solar image in this article. Therefore, the observing frequency of the solar images shown in this article is the same as the frequency of the first LO: 100 GHz for Band 3, 239 GHz for Band 6.

\begin{figure}[t]
\centering
\includegraphics[scale=.65]{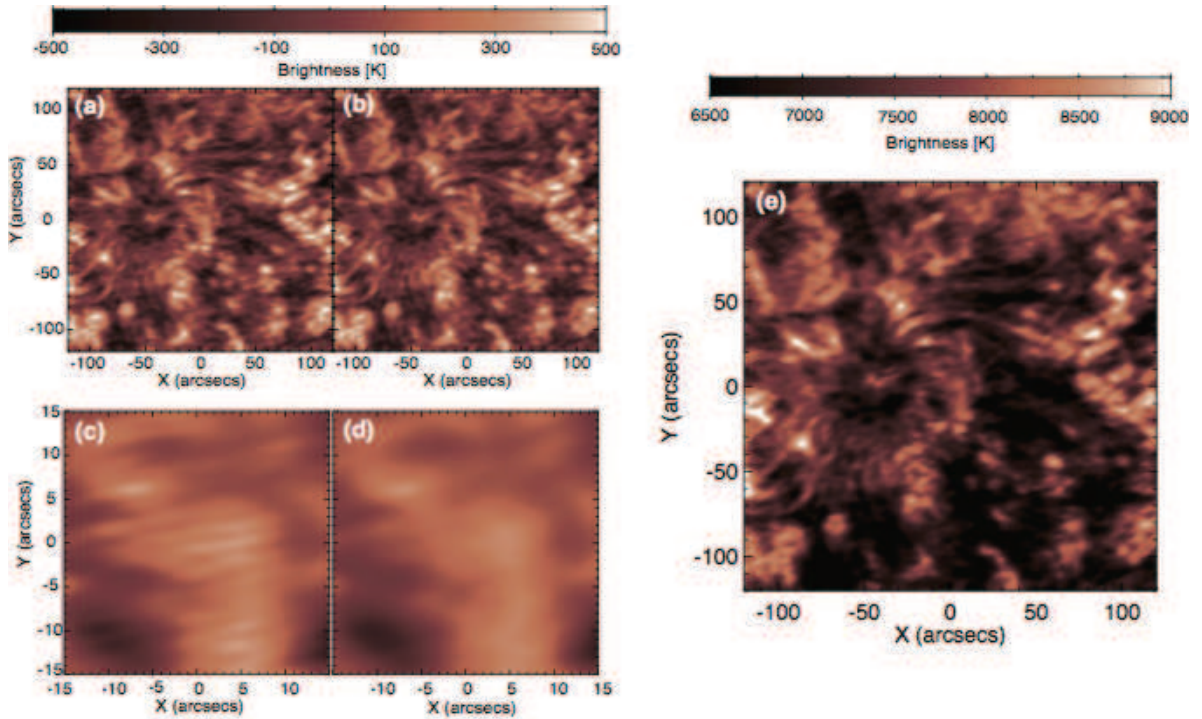}
\caption{The sunspot images synthesized from a 149-point mosaic observation with Band 3 using the MD2 mode. (a) and (b): Synthesized images of the lead sunspot of AR 12470; (c) and (d): Expanded images around the center of (a) and (b); (e): A combined image created from interferometric and single-dish observations. (a) and (c) are synthesized with a Briggs robust weighting factor \textsf{robust=0.0} (the CASA default). Images (b), (d), and (e) are synthesized with \textsf{robust=1.0}}
\end{figure}

Figure 7 shows the images of the leading sunspot in AR 12470 on 16 December 2015 synthesized from the 149-point mosaic observation in Band 3 (example 1 in Table~4). The default visibility weighting option of the CASA {\sf clean} task is to set the Briggs robust weighting parameter \citep{Briggs95} to zero. We note that artifacts appear in the image in the form of fine stripes, as evident in panels a and c. We attribute this to undue weight being given to longer interferometric baselines. In particular the locations of the centers of 12-m array and 7-m array are not the same; the distance between them is about 200m. Hence, data on baselines longer than 200 m are always included in solar data, even when the observation is done with the most compact configuration of the 12-m array, as was the case in December 2015. The resulting baseline distribution is non-optimum and the (nonlinear) image deconvolution process is subject to instability.  The weighting of these longer baselines can be reduced by applying more nearly ``natural weighting" \citep[\eg][]{2001isra.book.....T} by setting the Briggs robust weighting parameter to unity. When this is done, the artificial stripes disappear in the image (panels b and d of Figure 7). Hence, in this article, we always set the robust parameter to 1.0. The value is not fully optimized, and the most suitable value might depends on the target and array configuration. The angular resolution of the images shown in Figure 7b, d  -- \ie the dimension of the synthesized beam - is 4.9\arcsec$\times$2.2\arcsec.

The synthesized solar images include pixels with negative values. The negative values have physical meaning, because the interferometric data does not include the DC component of the brightness distribution in the field of view. Therefore, simultaneous single-dish observations are essential for obtaining absolute brightness temperatures from ALMA data. Figure 7e is the result of combining the synthesized image and the full-Sun map constructed from the simultaneous single-dish mapping data. The full-Sun map is created with CASA using the reduction \& imaging script included in the SV-data package. \citet{White17} pointed out that a correction factor has to be applied to any map created with CASA. The factor is applied to the full-Sun map used for creating the combined image shown in Figure 7e. We note that the correction factor is not applied to the full-Sun images of the SV data released on 18 January 2017.
 
For the combination, we use the default parameters of the {\sf feather} task in CASA. We found that the averaged brightness temperature of the combined image is always larger (5$\ --\ $10\%) than the temperature brightness at the same position in the single-dish map even though the values should be similar. This means that the parameters of the {\sf feather} task will need to be tuned in order to obtain consistent images, before using combined images for precise discussion of the absolute Tb.

\begin{figure}[t]
\centering
\includegraphics[scale=.9]{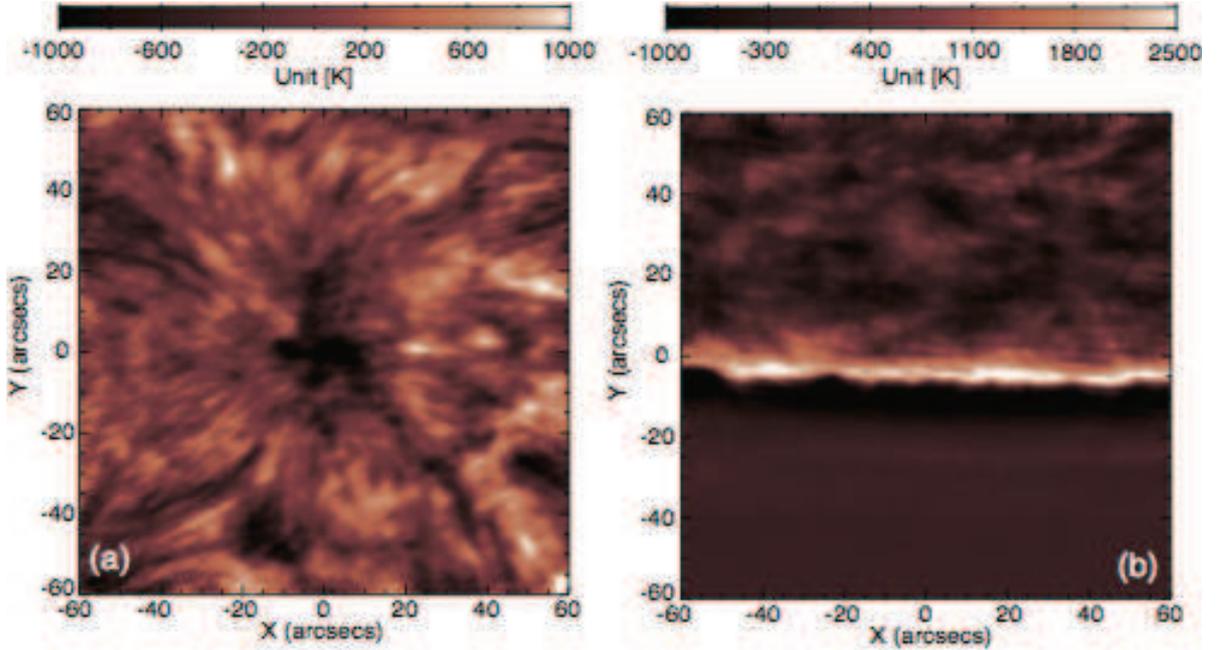}
\caption{The solar images synthesized from the 149-points mosaic observations with Band 6 using the MD2 mode. (a) The leading sunspot of AR 12470, (b) The solar limb around the South Pole. }
\end{figure}

Figure 8 presents solar images synthesized from a 149-point mosaic observation in Band 6 using the MD2 mode. Panel a shows the leading sunspot in AR 12470 (example 2 in Table~4) on 18 December 2015, and the panel b shows the solar limb near the South Pole (example 3 in Table~4). The calibration and synthesis imaging process are the same as those employed for Band 3, except for the observing frequency. Note that here the single dish data was not used for the images in Figure 8. The synthesized beams are $2.4\arcsec\times 0.9\arcsec$ for the sunspot image and $1.7\arcsec\times 1.0\arcsec$ ~deg for the South Pole image. We note that the narrow bright limb seen in the Figure 8b does not indicate ``limb brightening" that can be seen in a direct full-Sun image with radio. The value in a synthesized image instead indicates the derivation from the average brightness of the FoV that is determined mainly by the beam shape of an antenna ($\approx$25\arcsec\ at 239 GHz), even when we observe the target in the MOSAIC mode. The deviation at the solar disk near the limb appears anomalously large, because the brightness changes suddenly from the quiet sun level to the sky level. Thus, such a narrow bright limb appears only in the synthesized image. It should not be present in a combined image that is created from the synthesized image and full-Sun map.

\subsection{Estimating the Noise Level of Solar Synthesized images}

The noise level of a synthesized image may be determined from the \textsf{rms} value of the brightness on blank sky. However, this method cannot be applied to solar synthesized images because the primary beam of ALMA antennas is significantly smaller than the Sun in all frequency bands. Solar emission therefore completely fills the field of view in most cases, complicating the task of estimating noise. We therefore use an alternate method. ALMA is designed to support full polarimetry. To measure the Stokes-polarization parameters, the Band 3 and Band 6 receiver cartridges contain two complete receiver systems sensitive to orthogonal linear polarizations \citep{ALMAC4Tech}. We call one polarization X and the other one Y. The 64-input baseline correlator enables us to calculate four cross-correlations (XX, YY, XY, and YX) from the X- and Y-signals for each antenna baseline. However, only XX- and YY-cross-correlations are useful for solar observations in Cycle 4 because ALMA support of full Stokes polarimetry is not yet offered as a scientific capability. Nevertheless, we can synthesize images using XX- and YY-data that are observed simultaneously. In the absence of any flare emission, as was the case for the examples presented, solar mm/sub-mm emission is thermal emission from optically thick plasma \citep{1985ARA&A..23..169D}. Although there is possibility that the thermal emission is circularly polarized due to the presence of strong magnetic fields \citep{2004ASSL..314...71G,2016ApJ...818....8M,Loukitcheva17}, net linear polarization should be absent due to differential Faraday rotation, and we can assume that any such polarization at 100 GHz and 239 GHz is negligibly small in comparison with the precision of current ALMA solar observations. The crosstalk of the polarizations in the receiver system can be also neglected \citep{2006stt..conf..154C}. Therefore, the difference between the solar images synthesized from XX- and YY- data should be zero in principle, and the difference between the two polarizations can therefore be used as a proxy for the noise level in the final images (see Appendix A). 

\begin{figure}[t]
\centering
\includegraphics[scale=.8]{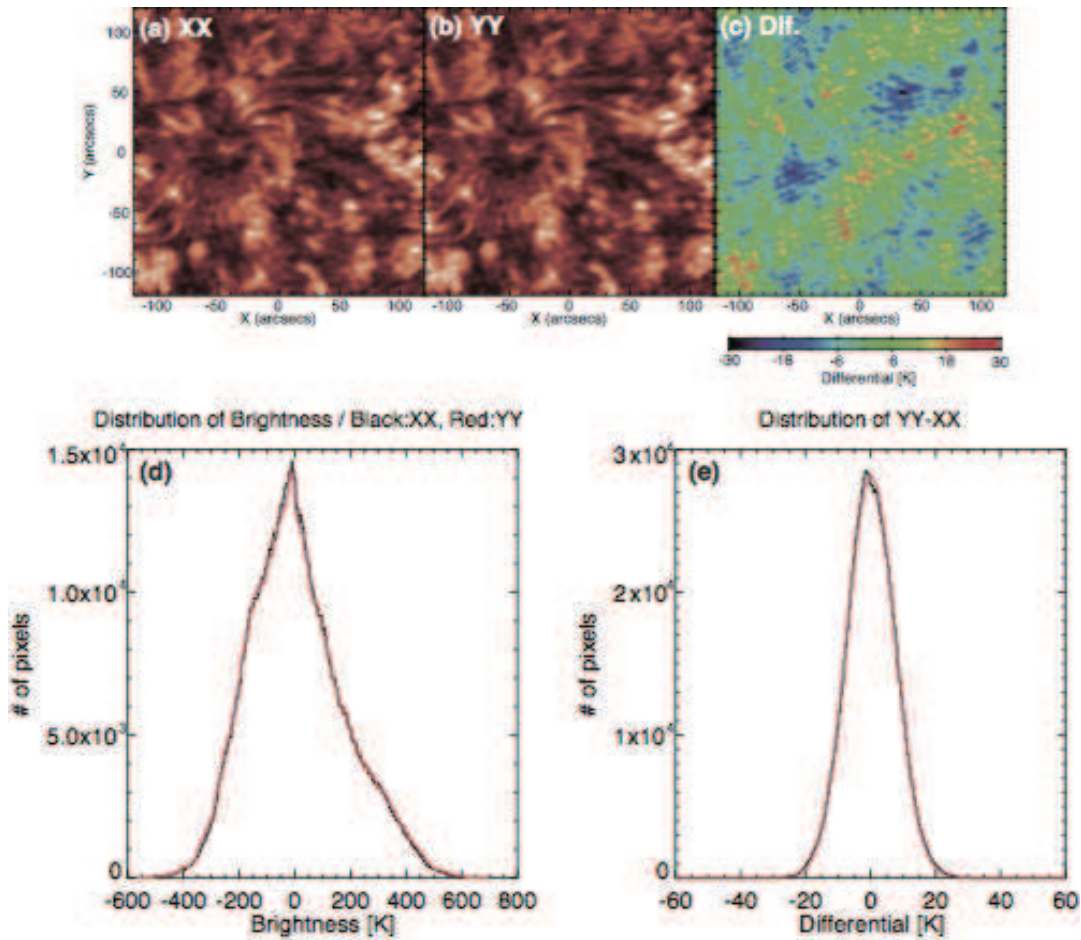}
\caption{(a) and (b) The Band 3 sunspot images synthesized from the data of XX and YY respectively; (c) The difference image of (a) and (b); (d) The pixel distributions of brightness in (a) [Black] and (b) [Red]; (e) the pixel distribution function of the difference image (c). The red line on (e) indicates the Gaussian function fit to the distribution.}
\end{figure}

Figure 9 shows estimations of the noise-level from maps formed using the XX- and YY-correlations. From the width of the Gaussian function fitted to the distribution of the differential (Figure 9e), the noise level of the Band 3 synthesized image of the sunspot (Figure 7b) is 3.7 K when the integration time is six seconds and the integration bandwidth is 8 GHz. We also apply the method to the sunspot image observed with Band 6 (Figure 8a), and estimate the noise level to be 9.8 K. The integration time and bandwidth of the Band 6 image are the same as those of the Band 3 image. 

\subsection{Imaging Artifacts Above the Solar Limb}

In addition to thermal noise, imaging artifacts may be present in a synthesis image as a result of incomplete sampling of the \uv-plane, non-optimum weighting of the visibility data (cf. Section 4.1), source variability, or other factors. An example of an artifact resulting from incomplete sampling and possibly non-optimum weighting is shown in Figure 10, in which a detail of the mosaic image of the South Pole is shown. Figure 10a shows a map made using the heterogeneous array comprised of 7 m antennas and 12 m antennas, as also shown in Figure 8b. Figure 10b shows the same image using only the 12 m antennas and Figure 10c shows the same image using only 7 m antennas. A stripe of negative flux density appears above the limb in Figure 10a and a stripe of positive flux density is seen even higher above the limb. The stripes are non-physical artifacts due to incomplete sampling of the ``step function" represented by the bright solar disk falling off to cold sky. The interferometric array shows a ``ringing" or ``overshoot" response as a result. In the image synthesized from only 7m antennas the positive enhancement is very weak (Figure 10c and the blue lines in Figure 10d, e) although the negative stripe persists. The image synthesized from only 12m antennas shows a stronger enhancement with a peak located about 20\arcsec\ above the limb (Panel b and the red lines in panels d and e of Figure 10). We note that the shortest baseline of the 12-m array observing the solar limb is 12.9 m, so the largest angular scale measured is 20.1\arcsec\ at 239 GHz. For the heterogeneous array the shortest baseline measured is 7.6 m, corresponding to an angular scale of 34.3\arcsec\ at 239 GHz. In principle, inclusion of the 7 m antennas should bridge the gap between the single-dish total-power map (resolution 24.4\arcsec) and the largest angular scale measured by the 12-m array. 

\begin{figure}[t]
\centering
\includegraphics[scale=.65]{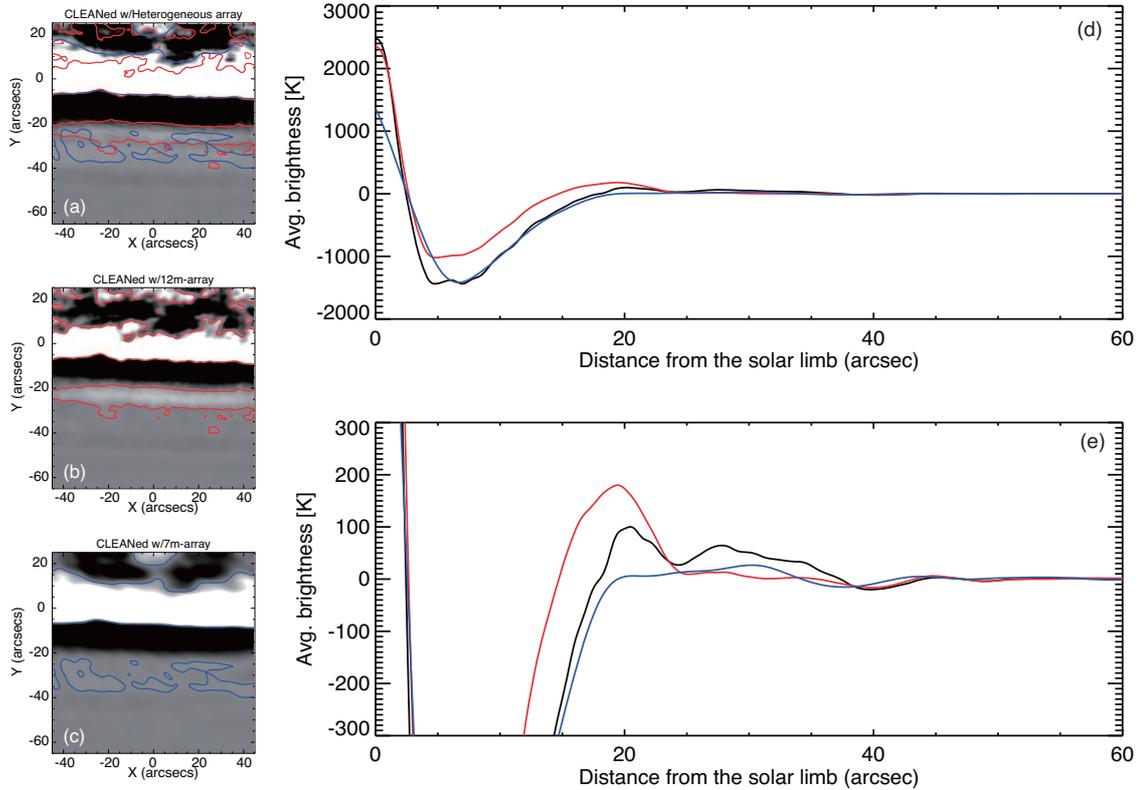}
\caption{The solar-limb images synthesized from the data of the (a) heterogeneous array, (b) 12-m array, and (c) 7-m array. The red and blue contours in the panels indicate +20 K level of 12-m array and 7-m array respectively. (d) and (e) show the brightness profiles as a function of the distance from the solar limb. Black: Heterogeneous array, Red: 12-m array, Blue: 7-m array. The difference of (d) and (e) is the range of {\it y}-axis.}
\end{figure}

A possible problem is mis-matched cross-calibration between 7 m and 12 m antennas. CASA currently supports two approaches to calibrating visibilities obtained with a heterogeneous array. In one, the data are jointly calibrated and in the other the data obtained with the 12-m array and 7-m array are calibrated independently and then combined.  We carried out the calibration of the data using both methods, and compared the resulting images. However, we cannot find any significant difference. Another possibility is that the relative weighting of the visibility baselines is incorrect: a careful assessment of the weights assigned to 7 m--7 m, 7 m--12 m, and 12 m--12 m baselines, as well as the weight given to the single dish total power map is needed. A final possibility is insufficient numbers of short antenna baselines. The 7-m array provides short baselines, and the visibilities of the baselines should suppress the sidelobes created by the 12-m array. In our case, we can see the suppression of the sidelobe by 7 m antennas (see the difference of the red and black lines in Figure 10). The remaining enhancement in the image synthesized from the data with the 7 m + 12 m heterogeneous array might indicate the lack of the short baselines. The commissioning observation is carried out using 9 $\times$ 7 m antennas and 21 $\times$ 12 m antennas. The number of the antennas is smaller than that for Cycle 4 observations and so there will be opportunities to better understand and resolve the issue. 

\subsection{Co-alignment between ALMA and other instruments}

\begin{figure}[t]
\centering
\includegraphics[scale=.60]{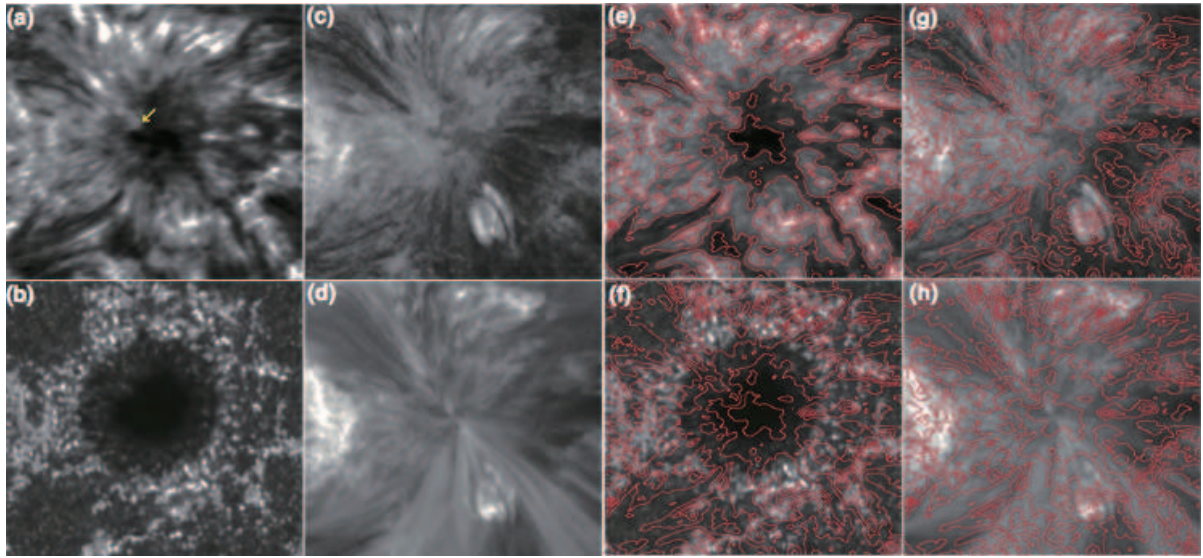}
\caption{Co-alignment between the ALMA Band 6 image and SDO/AIA images. The gray scale: (a) and (e): ALMA Band 6 synthesized image with the feathering process, (b) and (f) 1700 \AA \ band of AIA, (c) \& (g) 304 \AA \ band of AIA, (d) and (h) 193 \AA \ band of AIA. The red contours on (e), (f), (g), and (h) indicate the brightness of the ALMA Band 6 image a.}
\end{figure}

To maximize the scientific impact of ALMA data, it is very important to compare ALMA images with those obtained by instruments operating at other wavelengths with similar angular resolution. Direct comparisons require that ALMA images are accurately co-aligned with those produced by other instruments. ALMA operates in a geocentric coordinate frame using Right Ascension and Declination while heliocentric coordinate are usually used for solar imaging data. Therefore, ALMA images must be converted from RA/Dec coordinates to a heliocentric coordinate frame. 

The precision of the absolute pointing of the ALMA antennas is better than 2\arcsec\ \citep{ALMAC4Tech}. Figure 11 shows the result of co-alignment between the sunspot image with Band 6, UV continuum, and EUV images obtained with Solar Dynamics Observatory/Atmospheric Imaging Assembly \citep[SDO/AIA:][]{2012SoPh..275...17L}. For the co-alignments, we do not make any adjustment except for the coordinate conversion. It is hard to verify the co-alignment rigorously, because it is hard to find counterparts of the Band 6 images in the AIA images. The bright structure above the remnant of the light bridge in the AIA 304 \AA \ image is very similar to that in the Band 6 image. In comparing the edge of the structure in the umbra (yellow arrow in Figure 11a) the precision of the co-alignment appears to be better than the size of the synthesized beam (Figure 10e, g).

\begin{figure}[t]
\centering
\includegraphics[scale=.75]{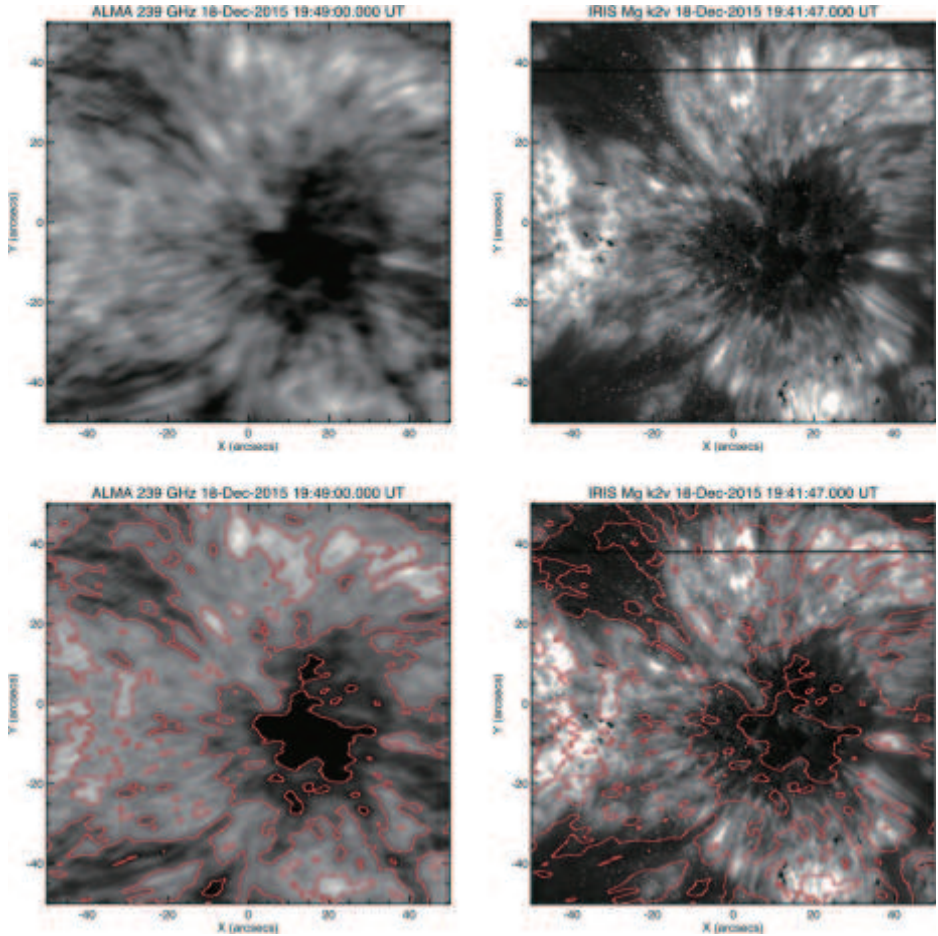}
\caption{Co-alignment between the ALMA Band 6 image and Mg k$_{\rm 2v}$ line image obtained with IRIS. Left: ALMA 239 GHz images. Right: Mg k$_{\rm 2v}$ image. Red contours indicate the brightness of the ALMA image.}
\end{figure}

Similarly, Figure 12 shows the result of the co-alignment between the Band 6 image and a Mg {\sc ii} k$_{\rm 2v}$ image obtained with the Interface Region Imaging Spectrograph \citep[IRIS:][]{2014SoPh..289.2733D}. In this case, we can easily identify the counterparts of the Band 6 image in the IRIS image. Therefore, the co-alignment is done only by the visual inspection. The similarity between the images suggests that Band 6 and Mg {\sc ii} k$_{\rm 2v}$ line emissions are formed within approximately the same range of heights.

\section{Concluding Remarks}

To conclude, this article summarizes the development and science-verification efforts leading up to the release of solar-observing modes by ALMA for Cycle 4 in 2016\ --\ 2017. While current capabilities remain limited, they represent a major advance over observational capabilities previously available at mm/submm wavelengths. Coupled with exciting space-based observations obtained by, \eg {\sl Hinode}, SDO, and IRIS; and ground based observations at, \eg National Solar Observatory, Big Bear Solar Observatory, Tenerife, and La Palma, ALMA opens a new window on contemporary scientific problems in solar physics. 

Current ALMA capabilities are summarized at the beginning of Section 4. Looking forward, additional capabilities are planned in support of solar observing that will greatly expand ALMA's science capabilities. It is planned that the following new capabilities will be available in the near future:

\begin{itemize}
\item Band 7 (275\ --\ 373 GHz: 850 $\mu$m) and Band 9 (602\ --\ 720 GHz: 450 $\mu$m) continuum observations of the Sun will be supported, in addition to Bands 3 and 6
\item Low-resolution spectroscopy (TDM mode) in Bands 3, 6, 7, and 9
\item Support of full Stokes polarimetry
\item Support of sub-second integration times
\end{itemize}

\noindent In the longer term, additional ALMA frequency bands will become available for use by the solar community. A number of other capabilities are under consideration, but the timing of their availability has not yet been established. These include the use of science subarrays, where the ensemble of 66 ALMA antennas can be divided into two or more independent arrays to perform multi-band or multi-target observations; band switching observations where an observer can change frequency bands on short times scales; fast-scan single dish mapping of small regions of the Sun -- \eg an active region -- on short time scales (tens of seconds); larger mosaics to enable imaging of larger regions on the Sun. The solar community will be informed about new capabilities for solar observing when the call for proposals is issued by the Joint ALMA Observatory each year. 

%
\begin{acks}

The ALMA solar commissioning effort was supported by ALMA Development grants from NAOJ (for the East Asia contribution), NRAO (for the North American contribution), and ESO (for the European contribution). The help and cooperation of engineers, telescope operators, astronomers--on--duty, Extension and Optimization of Capabilities (EOC; Formerly Commissioning and Science Verification) team, and staff at the ALMA Operations Support Facility was crucial for the success of solar commissioning campaigns in 2014 and 2015. We are grateful to the ALMA project for making solar observing with ALMA possible. This article makes use of the following ALMA data: ADS/JAO.ALMA\#2011.0.00020.SV, ADS/JAO.ALMA\#2011.0.00001.CAL. ALMA is a partnership of ESO (representing its member states), NSF (USA), NINS (Japan), together with NRC (Canada), NSC, ASIAA (Taiwan), and KASI (Republic of Korea), in cooperation with the Republic of Chile. The Joint ALMA Observatory is operated by ESO, AUI/NRAO and NAOJ. The National Radio Astronomy Observatory is a facility of the National Science Foundation operated under cooperative agreement by Associated Universities, Inc. SDO is the first mission to be launched for NASA's Living With a Star (LWS) Program. IRIS is a NASA small explorer mission developed and operated by LMSAL with mission operations executed at NASA Ames Research center and major contributions to downlink communications funded by ESA and the Norwegian Space Centre. This work was partly carried out on the solar data analysis system and common-use data-analysis computer system operated by Astronomy Data Center of NAOJ. M. Shimojo was supported by JSPS KAKENHI Grant Number JP17K05397. R. Braj\v{s}a acknowledges partial support of this work by Croatian Science Foundation under the project 6212 Solar and Stellar Variability and by the European Commission FP7 project SOLARNET (312495, 2013 - 2017), which is an Integrated Infrastructure Initiative (I3) supported by the FP7 Capacities Programme. G.D. Fleishiman acknowledges support from NSF grants AGS-1250374 and AGS-1262772. The trip of Y. Yan to 2015 ALMA Solar Campaign was partially supported by NSFC grant 11433006. S. Wedemeyer acknowledges funding from the European Research Council (ERC) under the European Union's Horizon 2020 research and innovation programme (grant agreement No 682462)

\end{acks}

\appendix   

ALMA antennas measure the two orthogonal linear polarizations X and Y, and the 64-input baseline correlator measures the products of the linearly polarized antenna voltages. For a pair of antennas, $m$ and $n$, the correlation products are $v_{\rm x_mx_n}$, $v_{\rm y_my_n}$, $v_{\rm x_my_n}$, and $v_{\rm y_mx_n}$. For well-designed antenna feeds and weakly polarized emission \citep{1999ASPC..180..111C}, the response of the interferometer can be expressed as 
\begin{eqnarray*}
v'_{\rm xx} &=& g_{\rm mx} g^\ast_{\rm nx} (I + Q \cos 2\chi + U \cos 2\chi) + \sigma'_{\rm xx} \\
v'_{\rm xy} &=& g_{\rm mx} g^\ast_{\rm ny} ((d_{\rm mx}-d^\ast_{\rm ny})I - Q \cos 2\chi + U \cos 2\chi + jV) + \sigma'_{\rm xy} \\
v'_{\rm yx} &=& g_{\rm my} g^\ast_{\rm nx} ((d^\ast_{\rm nx}-d_{\rm my})I - Q \cos 2\chi + U \cos 2\chi - jV) + \sigma'_{\rm yx} \\
v'_{\rm yy} &=& g_{\rm my} g^\ast_{\rm ny} (I - Q \cos 2\chi - U \cos 2\chi) + \sigma'_{\rm yy}
\end{eqnarray*}
\noindent where $I$ is Stokes parameter describing the total intensity of the radiation, $Q$ and $U$ are the Stokes parameters characterizing linearly polarized radiation, and $V$ is the Stokes parameter characterizing circularly polarized radiation. The parallactic angle [$\chi$] includes the effects of rotation of the alt--az ALMA antennas as viewed from the source. The $g$-factors are complex gain factors established by calibration and the $d$-terms represent polarization ``leakage" which, by careful design, are small but measurable complex numbers, also determined by calibration. The noise in each correlation measurement is represented by $\sigma'$. At present ALMA does not support full Stokes polarimetry and in particular, measurements of Stokes-$V$, which requires calibration of the complex leakage terms. However, it is expected that support of full Stokes polarimetry will be implemented soon, thereby enabling a powerful new probe of chromospheric magnetic fields. For the present purpose, however, only the parallel correlations are of interest here. Rearranging $v_{xx}$ and $v_{yy}$ we have 
\begin{eqnarray*}
(I + Q \cos 2\chi + U \cos 2\chi) &=& v'_{\rm xx}/(g_{\rm mx} g^\ast_{\rm nx}) + \sigma_{\rm xx}'/(g_{\rm mx} g^\ast_{\rm nx}) = v_{\rm xx} + \sigma_{\rm xx}\\
(I - Q \cos 2\chi - U \cos 2\chi) &=& v'_{\rm yy}/(g_{\rm my} g^\ast_{\rm ny}) + \sigma_{\rm yy}'/(g_{\rm my} g^\ast_{\rm ny}) = v_{\rm yy} + \sigma_{\rm yy}
\end{eqnarray*}
\noindent where the unprimed quantities represent calibrated measurements. Summing and differencing these quantities and propagating the noise terms yields
\begin{eqnarray*}
I &=& {1\over 2}(v_{\rm xx} + v_{\rm yy}) + \sigma_{\rm I} \\
Q \cos 2\chi &+& U \cos 2\chi = {1\over 2}(v_{\rm xx} - v_{\rm yy})+ \sigma_{\rm I}
\end{eqnarray*}
\noindent where $\sigma_{\rm I}=\sqrt{\sigma^2_{\rm xx}+\sigma^2_{\rm yy}}/2$. It is seen that the sum of the calibrated correlation products $v_{\rm xx}$  and $v_{\rm yy}$ for a given antenna pair represents the interferometer's response to Stokes $I$. While the Stokes-$V$ parameter may be non-zero the Stokes-$Q$ and $U$ parameters are expected to be zero for thermal solar emission and so $(v_{\rm xx} - v_{\rm yy})/2 =  \sigma_{\rm I}$. Note further that for emission that is not linearly polarized, the calibrated noise terms are such that $\sigma_{\rm xx}=\sigma_{\rm yy}$ and so $\sigma_{\rm I}=\sigma_{\rm xx}/\sqrt{2} = \sigma_{\rm yy}/\sqrt{2}$.  Since synthesis maps represent a linear superposition of interferometric measurements, the same relation holds true for synthesis images.

%
%

%

\end{article} 
\end{document}